\documentclass[aps,nofootinbib,showpacs,preprintnumbers,amsmath,amssymb]{revtex4}
\def\be{\begin{equation}}
\def\ee{\end{equation}}
\def\bea{\begin{eqnarray}}
\def\eea{\end{eqnarray}}
\newcommand{\A}{{\mathcal{A}}}
\newcommand{\tA}{{\widetilde {\mathcal{A}}}}
\newcommand{\ta}{{\widetilde a}}
\newcommand{\tal}{{\widetilde \alpha}}
\newcommand{\tu}{{\widetilde u}}

\newcommand{\td}{{\widetilde d}}
\newcommand{\tQ}{{\widetilde Q}}

\newcommand{\MSbar}{\overline{\rm MS}}

\usepackage{graphics}
\usepackage{graphicx}
\usepackage{dcolumn}
\usepackage{bm}
\usepackage{epsfig}
\usepackage{graphicx}
\usepackage{multirow}

\begin{document}

\vspace{1cm}

\preprint{USM-TH-305, arXiv:1209.2953v3}

\begin{flushright} {\bf} \end{flushright}
 
\title{Operator Product Expansion with analytic QCD in $\tau$ decay physics}
\footnote{
In comparison with v2: rewritten and extended Section II; 
Section III refers to a new Appendix A; 
the results in Section IV are obtained by the standard minimization 
involving $\chi^2$;
Conclusions include a comparison with other results
in the literature for the gluon condensate; new references [47] and [58].
To appear in PRD.} 

\author{Gorazd Cveti\v{c}$^{1}$}
\author{Cristi\'an Villavicencio$^{1,2}$}
\affiliation{$^1$Department of Physics and Centro Cient\'{\i}fico-Tecnol\'ogico de Valpara\'{\i}so,\\
Universidad T{\'e}cnica Federico Santa Mar{\'\i}a, 
Casilla 110-V, Valpara{\'\i}so, Chile\\
$^2$Universidad Diego Portales, Casilla 298-V, Santiago, Chile}

\date{\today}

\begin{abstract}
We apply a recently constructed model of analytic QCD in the
Operator Product Expansion (OPE) analysis of the $\tau$ lepton decay data
in the $V$+$A$ channel. The model has the running coupling $\A_1(Q^2)$
with no unphysical singularities, i.e., it is analytic. 
It differs from the corresponding perturbative QCD coupling $a(Q^2)$ 
at high squared momenta $|Q^2|$ by terms $\propto (1/Q^2)^5$, 
hence it does not contradict the ITEP OPE philosophy and can be 
consistently applied with OPE up to terms of dimension $D=8$.
In evaluations for the Adler function we use a 
Pad\'e-related renormalization-scale-independent resummation, 
applicable in any analytic QCD model. Applying the Borel sum
rules in the $Q^2$ plane along rays 
of the complex Borel scale and comparing with ALEPH data of 1998, 
we obtain the gluon condensate value 
$\langle (\alpha_s/\pi)  G^2 \rangle = 0.0055 \pm 0.0047 \ {\rm GeV}^4$.
Consideration of the $D=6$ term gives us the result 
$\langle O_6^{(V+A)} \rangle = (-0.5 \pm 1.1) \times 10^{-3} \ {\rm GeV}^6$,
not incompatible with nonnegative values. 
The real Borel transform gives us then,
for the central values of the two condensates, a good agreement with
the experimental results in the entire considered interval of 
the Borel scales $M^2$. 
In perturbative QCD in $\MSbar$ scheme we deduce 
similar result for the gluon condensate, 
$0.0059 \pm 0.0049 \ {\rm GeV}^4$, 
but the value of $D=6$ condensate is negative, 
$\langle O_6^{(V+A)} \rangle = (-1.8 \pm 0.9) \times  10^{-3} \ {\rm GeV}^6$, 
and the resulting
real Borel transform for the central values is close to the lower
bound of the experimental band.
\end{abstract}

\maketitle

\section{Introduction}
\label{sec:intr}
Perturbative QCD (pQCD) in the usual renormalization schemes, such as
$\MSbar$, has the peculiar property of resulting in a running
coupling $a(Q^2)$ ($\equiv \alpha_s(Q^2)/\pi$) which has singularities
in the complex $Q^2$ plane outside of the negative semiaxis, often
at positive $Q^2 \equiv - q^2 >0$. This implies that
spacelike physical  quantities ${\cal D}(Q^2)$ evaluated in terms of $a(Q^2)$ have
the same type of singularities, contravening the general principles 
of (causal and local) quantum field theories \cite{BS,Oehme}.
These problematic aspects of pQCD were addressed in the seminal
works of Shirkov, Solovtsov et al.~\cite{ShS,MSS,Sh} who constructed, as
an alternative, a QCD model which has no such unphysical singularities.
Their idea was to keep the discontinuity function
$\rho_1^{\rm (pt)}(\sigma) = {\rm Im} \ a(-\sigma - i \epsilon)$ unchanged for 
$Q^2 \leq 0$ ($\sigma \geq 0$), but eliminating the offending singularities
at $Q^2>0$ by not including them in the dispersion relation for their
coupling. The same principle was applied to any other integer power
of $a(Q^2)$. This approach was called Analytic Perturbation Theory
(APT); in the present context we call it Minimal Analytic (MA) QCD,
due to the unchanged (perturbative) singularities along the 
$Q^2 \leq 0$ semiaxis. For various applications of MA and related models, see 
Refs.~\cite{Sh,MSSY,Milton:2001mq,mes2,Pase,NesSim,Bdecays,rev1,rev2,rev3}.\footnote{
For a somewhat related but different approach, which performs minimal
analytization of $d \ln a(Q^2)/d \ln Q^2$ (and not: $a(Q^2)$),
see Refs.~\cite{Nest1}, and for extensions thereof Refs.~\cite{Nest2}.} 
However, MA has two aspects which,
under specific circumstances, may be regarded as inconvenient: 
\begin{enumerate}
\item
It subestimates \cite{MSS,MSSY}
the semihadronic $\tau$ decay ratio\footnote{
$r_{\tau}$ represents the QCD part of the strangeless $V$+$A$
decay ratio $R_{\tau}(\Delta S=0)$; it
has the (small) quark mass effects subtracted and is
normalized in the canonical way: $r_{\tau} = a + {\cal O}(a^2)$.
In Ref.~\cite{Milton:2001mq} the correct value of $r_{\tau}$ is reproduced in
MA at a price of (modifying MA by) introducing strong mass threshold
effects in the low-momentum regime.}
$r_{\tau}$, giving $r_{\tau}({\rm MA}) \approx 0.13$-$0.14$ in the
strangeless $V$+$A$ channel while the experimental  
value is $0.203 \pm 0.004$ \cite{ALEPH2,DDHMZ}. 
\item
At momenta
$Q^2 > \Lambda^2$ (where $\Lambda^2 \sim 0.1 \ {\rm GeV}^2$), the
MA coupling $\A_1^{\rm (MA)}(Q^2)$ differs from the underlying
QCD coupling $a(Q^2)$ by terms $\sim (\Lambda^2/Q^2)$.
The latter property implies that, in MA, the leading-twist
contributions to physical quantities contain power terms
$\sim (\Lambda^2/Q^2)$ which are of ultraviolet origin. This is
not in accordance with the ITEP interpretation of OPE
\cite{Shifman:1978bx,DMW} which states that the higher-dimension
(higher-twist) terms $\sim (\Lambda^2/Q^2)^n$ in OPE are of infrared origin. 
\end{enumerate}

These problems have been addressed in a series of papers
\cite{panQCD,1danQCD,2danQCD}. In Ref.~\cite{panQCD} the
following question was investigated: Is there an analytic QCD
which is simultaneously fully perturbative and reproduces the
correct value of $r_{\tau}$? The results of  Ref.~\cite{panQCD}
indicate that there is probably no such framework, unless we
introduce renormalization schemes which make the perturbation
series highly divergent starting at terms $\sim a^5$. Therefore,
in Refs.~\cite{1danQCD,2danQCD} a more modest goal was pursued:
construction of analytic QCD which is not fully perturbative, but
addresses the previously mentioned two points nonetheless: the
correct value of $r_{\tau}$ is reproduced, and the deviation
of the analytic coupling $\A_1(Q^2)$ from the 
underlying perturbative coupling $a(Q^2)$ at high $|Q^2|$ 
is proportional to a ``sufficiently high'' 
power of $1/Q^2$. In Ref.~\cite{1danQCD}, an analytic
model was constructed for which 
$\A_1(Q^2) - a(Q^2) \sim (\Lambda^2/Q^2)^3$ at high $|Q^2| > \Lambda^2$.
This was achieved by constructing the model-defining 
discontinuity function 
$\rho_1(\sigma) \equiv {\rm Im} \ \A_1(-\sigma - i \epsilon)$ in such a way that
it is equal to the discontinuity function 
$\rho_1^{\rm (pt)}(\sigma) \equiv {\rm Im} \ a(-\sigma - i \epsilon)$ of
the underlying pQCD coupling $a(Q^2)$ at $\sigma > M_0^2$ (where
$M_0 \sim 1$ GeV), and in the low-$\sigma$ regime 
$(0< \sigma < M_0^2)$ the unknown behavior of $\rho_1(\sigma)$ was
parametrized by a single delta function. By adjusting the
free parameters of the model, the aforementioned suppressed
deviation $\A_1(Q^2) - a(Q^2) \sim (\Lambda^2/Q^2)^3$ at $Q^2 > \Lambda^2$
was achieved.\footnote{
This condition was also achieved 
in the analytic model of Ref.~\cite{Alekseev}.}
In Ref.~\cite{2danQCD}, this idea was continued,
by employing in the low-energy regime ($0 < \sigma < M_0^2$)
the parametrization in terms of two delta functions for the
discontinuity function $\rho_1(\sigma)$. In this way, the
strongly suppressed deviation  $\A_1(Q^2) - a(Q^2) \sim (\Lambda^2/Q^2)^5$ 
at $Q^2 > \Lambda^2$ was achieved (while at the same time
the correct value of $r_{\tau}$ was reproduced). Therefore,
in this model the leading-twist contributions to physical 
quantities do not give power-suppressed terms of
dimension $D < 10$ of ultraviolet origin. Hence, the model
can be applied with OPE, without contradicting the ITEP interpretation 
of OPE, up to $D=8$ terms.

We wish to point out that our approach eliminates unphysical singularities
from the QCD coupling while still preserving the salient features of the
usual pQCD approach, among them the applicability of OPE (in the ITEP sense)
and the related universality of the QCD coupling. 
On the other hand, the approaches 
of Refs.~\cite{Milton:2001mq,mes2,Nest3,DeRafael,MagrDual} follow a different
line: by considering (various) specific timelike observables,
they eliminate the unphysical singularities directly from the
corresponding spacelike observables. This leads either directly 
\cite{Milton:2001mq,mes2,Nest3} or indirectly \cite{DeRafael,MagrDual}
to analytic QCD couplings whose nonunversality is reflected in
observable-dependent modifications in the low-energy
(low-$\sigma$) regime.

In this work, we will apply the mentioned analytic model
of Ref.~\cite{2danQCD} to the analysis of the $\tau$-decay data,
in the $V$+$A$ channel, using the OPE approach (with up to $D=6$ terms)
with Borel sum rules,
along the lines of Refs.~\cite{Geshkenbein,Ioffe} where the
analysis was performed within the perturbative QCD in $\MSbar$ scheme.
The program of the work is the following.
In Sec.~\ref{sec:model} we summarize the previously
mentioned QCD analytic model \cite{2danQCD}.
In Sec.~\ref{sec:resum} we recapitulate a powerful 
Pad\'e-related resummation method \cite{BGA}
which is very natural and convenient 
to apply in evaluations of physical quantities within analytic QCD 
models. In Sec.~\ref{sec:Borel} we apply this resummation, 
within our model, in the evaluation of the (leading-dimension part 
of the) Adler function, i.e., 
the logarithmic derivative of the polarization operator $\Pi(Q^2)$ 
in the massless limit. We combine this evaluation, and
the OPE expansion of the Adler function, with the Borel sum rules 
along rays in the complex plane of the Borel scales $M^2$
(as in Refs.~\cite{Geshkenbein,Ioffe}; cf.~also.~Ref.~\cite{Ioffe:2000ns}).
For the experimental input we use ALEPH results of 1998 for 
$\omega_{\rm exp}(\sigma)$ ($\propto {\rm Im} \ \Pi(-\sigma - i \epsilon)$)
obtained from $\tau$-decay invariant-mass spectra for $0 < \sigma <
m_{\tau}^2$ \cite{ALEPH1}. We also discuss the obtained results
and compare them with the results obtained when no resummation is
performed in analytic QCD, and with the results in perturbative QCD
in the renormalization scheme that is underlying the analytic QCD model
(with and without resummation) and in $\MSbar$ renormalization scheme. 
Section \ref{sec:concl} 
contains a summary of the obtained results and conclusions.

\section{2-delta analytic QCD}
\label{sec:model}

Here we recapitulate the analytic QCD model of Ref.~\cite{2danQCD},
which contains a parametrization with two delta functions
in the low-$\sigma$ regime of the discontinuity function
$\rho_1(\sigma) = {\rm Im} \ \A_1(-\sigma - i \epsilon)$. 
Any analytic QCD model is defined, in principle, via its
analytic spacelike coupling $\A_1(Q^2)$, which is the analytic analog 
of the perturbative coupling $a(Q^2) \equiv \alpha_s(Q^2)/\pi$. Any
other coupling $\A_n(Q^2)$ (analog of the power $a(Q^2)^n$) can then
be constructed from it, via the formalism introduced in 
Refs.~\cite{GCCV} when $n$ is integer and in Ref.~\cite{GCAK} when
$n$ is general noninteger.\footnote{A formalism for construction of
$\A_n(Q^2)$ for general noninteger $n$, applicable only to MA,
was developed and applied in Refs.~\cite{BMS}.}
On the other hand, the coupling $\A_1(Q^2)$, analytic
in the complex plane outside of the negative semiaxis
$Q^2 \in \mathbb{C} \backslash (-\infty, 0]$, 
can be expressed via the discontinuity function 
$\rho_1(\sigma) \equiv {\rm Im} \ \A_1(-\sigma - i \epsilon)$
which is defined only for $\sigma >0$. The expression is a dispersion
relation which follows from the application of the Cauchy theorem,
the analyticity of $\A_1(Q^2)$, and the asymptotic freedom at
large $|Q^2|$ 
\be
\A_1(Q^2) 
= \frac{1}{\pi} \int_{0}^{+\infty} \ d \sigma 
\frac{ \rho_1(\sigma) }{(\sigma + Q^2)} \ .
\label{dispA1}
\ee
Due to the success of perturbative QCD at high energies, it is
reasonable to assume that at higher $\sigma$ ($\sigma > M_0^2$,
where $M_0 \agt 1$ GeV) the discontinuity function
$\rho_1(\sigma)$ coincides with the underlying perturbative
function $\rho_1^{\rm (pt)}(\sigma) \equiv {\rm Im} \ a(-\sigma - i \epsilon)$.
In the low-$\sigma$ regime ($0 < \sigma < M_0^2$) the behavior is
unknown in its details, and is here parametrized with two
delta functions\footnote{
\label{dispDA1}
We note that the function
\begin{displaymath}
\triangle \A_1(Q^2) \equiv \A_1(Q^2) -
\frac{1}{\pi} \int_{M_0^2}^{+\infty} \ d \sigma 
\frac{ \rho_1(\sigma) }{(\sigma + Q^2)} =
\frac{1}{\pi} 
\int_{M_{\rm thr}^2}^{M_0^2} \ d \sigma 
\frac{ \rho_1(\sigma) }{(\sigma + Q^2)} \ ,
\end{displaymath}
is a Stieltjes function (and $M_{\rm thr} \sim 10^{-1}$ GeV is a QCD threshold scale).
Approximating the discontinuity 
function $\rho(\sigma)$ of any Stieltjes function $f(Q^2)$ 
as a sum of delta functions is well motivated, 
cf.~Refs.~\cite{Peris,CM}, because it leads to
approximating the Stieltjes function by near-to-diagonal 
Pad\'e approximants. The latter must converge to the Stieltjes
function when the order (i.e., the number of deltas) increases
\cite{Pades}. An idea similar to Eq.~(\ref{rho1o1}),
but with one delta, was applied in Refs.~\cite{DeRafael,MagrDual}
directly to spectral functions of the vector current correlators.} 
at lower values $\sigma = M_1^2$ and $M_2^2$
($0 < M_2^2 < M_1^2 < M_0^2$)
\bea
\rho_1^{\rm (2 \delta)}(\sigma; c_2) &=&
\pi \sum_{j=1}^2 f_j^2 \Lambda^2 \; 
\delta(\sigma - M_j^2) +  \Theta(\sigma-M_0^2) \times 
\rho_1^{\rm (pt)}(\sigma; c_2) 
\label{rho1o1}
\\
& = & \pi \sum_{j=1}^2 f_j^2 \; \delta(s - s_j) +  
\Theta(s-s_0) \times r_1^{\rm (pt)}(s; c_2) \ .
\label{rho1o2}
\eea
In Eq.~(\ref{rho1o2}) dimensionless parameters were
introduced: $s=\sigma/\Lambda^2$, $s_j = M_j^2/\Lambda^2$ ($j=0,1,2$),
and $r_1^{\rm (pt)}(s; c_2) =  \rho_1^{\rm (pt)}(\sigma; c_2)
= {\rm Im} \ a(Q^2=-\sigma - i \epsilon; c_2)$, $c_2$ being a scheme parameter. 
The Lambert scale $\Lambda^2$ ($\stackrel{<}{\sim} 10^{-1} \ {\rm GeV}^2$)
will be defined below.
The underlying perturbative coupling $a(Q^2)$ is
taken, for convenience, in such renormalization schemes 
where it can be expressed explicitly as a solution
in terms of the Lambert function $W(z)$
\bea
a(Q^2;c_2) = - \frac{1}{c_1} \frac{1}{\left[
1 - c_2/c_1^2 + W_{\mp 1}(z) \right]} \ .
\label{aptexact}
\eea
Here, $Q^2=|Q^2| \exp(i \phi)$; $W_{-1}$ and $W_{+1}$
are the branches of the Lambert function
for $0 \leq \phi < + \pi$ and $- \pi < \phi < 0$, 
respectively,\footnote{
The functions $W_{\pm 1}(z)$ are implemented
in MATHEMATICA \cite{Math8} by the commands ${\rm ProductLog}[\pm,z]$.}
and the variable $z$ involves the mentioned Lambert scale
$\Lambda$
\be
z =  - \frac{1}{c_1 e} 
\left( \frac{|Q^2|}{\Lambda^2} \right)^{-\beta_0/c_1} 
\exp \left( - i {\beta_0}\phi/c_1 \right) \ .
\label{zexpr}
\ee 
The explicit expression (\ref{aptexact}) was presented
in Ref.~\cite{Gardi:1998qr}. It is the solution 
of the following (perturbative) renormalization group equation (RGE):
\bea
\frac{\partial a(Q^2;c_2)}{\partial \ln Q^2} & = &
- \beta_0 a^2 \frac{\left[ 1 + (c_1 - (c_2/c_1)) a
\right]}{\left[ 1 - (c_2/c_1) a \right]} \ .
\label{RGE}
\eea
Here, $\beta_0 = (1/4) (11 - 2 n_f/3)$ and 
$c_1 = \beta_1/\beta_0 = (1/4) (102-38 n_f/3)/(11 - 2 n_f/3)$
are universal constants. On the other hand, $c_2 \equiv \beta_2/\beta_0$ is the
free three-loop renormalization scheme parameter.
The expansion of the above beta function $\beta(a) =
\partial a/\partial \ln Q^2$ gives
\be
\beta(a) = 
- \beta_0 a^2 
(1 + c_1 a + c_2 a^2 + c_3 a^3 + \ldots ) \ ,
\label{betapt}
\ee
with the higher renormalization scheme parameters
$c_j \equiv \beta_j/\beta_0$ ($j\geq 3$) fixed by the value of $c_2$:  
$c_j = c_2^{j-1}/c_1^{j-2}$ ($j \geq 3$).\footnote{
Further use of Lambert functions in QCD running is discussed
also in Refs.~\cite{Kou,KuMa,Magr2,GCIK}.}

Application of the dispersion relation (\ref{dispA1}) to the
discontinuity function (\ref{rho1o2}) gives the analytic coupling
$\A_1(Q^2)$ of the model
\be
\A_1(Q^2;c_2) = \sum_{j=1}^2 \frac{f_j^2}{(u +s_j)} +
\frac{1}{\pi} \int_{s_0}^{\infty} ds \; 
\frac{r_1^{\rm (pt)}(s;c_2)}{(s+u)} \ ,
\label{A1Q2}
\ee
where $u = Q^2/\Lambda^2$. 
The free parameters of the model ($c_2$, $\Lambda$, $s_0$, $s_1$, $f_1^2$, 
$s_2$, $f_2^2$) are fixed by various requirements. The scale
$\Lambda$ is fixed by requiring that the underlying perturbative coupling
$a(Q^2)$ reproduce the central value of the world average
at $Q^2=M_Z^2$ as given in Ref.~\cite{PDG2010}:
$a^{({\overline {\rm MS}})}(M_Z^2) = 0.1184/\pi$.
More specifically, the coupling $a(Q^2;c_2)$, with a chosen value of $c_2$
and in the considered low-momentum regime of interest 
($Q < 2 {\overline m}_c$ where $n_f=3$), is first transformed
to the exact four-loop $\MSbar$ scheme, then RGE-evolved up to $Q^2=M_Z^2$ 
using the four-loop $\MSbar$ beta function
and at the quark thresholds $Q^2_{\rm thr} = (2 {\overline m}_q)^2$
using the three-loop matching
conditions \cite{CKS}. The Lambert scale $\Lambda$ (at $n_f=3$) is 
then fixed such that
the described procedure leads to the value 
$a^{({\overline {\rm MS}})}(M_Z^2) = 0.1184/\pi$.
The four parameters $s_j$ and $f_j^2$ 
($j=1, 2$), at chosen values of $s_0$ and $c_2$, are determined as 
functions of $s_0$ by the earlier mentioned (ITEP-OPE motivated)
requirement $\A_1(Q^2) - a(Q^2) \sim (\Lambda^2/Q^2)^5$ for $|Q^2| > \Lambda^2$
(these are four requirements, in fact). Subsequently, the value of the
parameter $s_0$ ($\equiv M_0^2/\Lambda^2$) is determined, 
for a chosen value of $c_2$,
by the requirement that the model reproduce the aforementioned
experimental value of the $V$+$A$ strangeless 
semihadronic $\tau$ decay ratio $r_{\tau} \approx 0.203$.
Finally, the values of the parameter $c_2$ are varied in such a way
that the resulting pQCD-onset scale $M_0$ ($\sim 1$ GeV) varies 
within a specific range.
For phenomenological reasons, the preferred values of
$M_0$ should be below and not too close to the value of the $\tau$ lepton
mass ($m_{\tau} = 1.777$ GeV). On the other hand, 
if $M_0 < 1.15$ GeV, the values
of $\A_1(Q^2)$ become too large ($\A_1(0) > 1$) 
and the model loses stability in the infrared.
We choose the representative values of the scheme parameter
$c_2$ such that the following conditions are fulfilled:
$1 \ {\rm GeV} < M_0 < 1.5 \ {\rm GeV}$ and $\A_1(0) \leq 1$.
For the central value $M_0=1.25$ GeV, we have the values of parameters 
given in the second line of Table \ref{t1}.
The two border choices ($M_0 = 1$ GeV; and $\A_1(0)=1$ with
$M_0 \approx 1.15$ GeV) are also given in Table \ref{t1}.\footnote{
In Ref.~\cite{2danQCD}, we included $M_0=1.00$ GeV.
However, in this case $\A_1(0) = 2.29$, indicating instability 
in the infrared.}
\begin{table}
\caption{Values of the parameters of the considered two-delta 
anQCD model, under the restriction for the values of the 
pQCD-onset scale $M_0 \equiv \sqrt{s_0} \Lambda$: 
$1 \ {\rm GeV} \leq M_0 \leq 1.5 \ {\rm GeV}$,
and $\A_1(0) \leq 1.0$.
The corresponding Lambert scales $\Lambda$ are for 
the central value of the QCD
coupling parameter $\alpha_s^{({\overline {\rm MS}})}(M_Z^2) = 0.1184$.}
\label{t1}  
\begin{ruledtabular}
\begin{tabular}{l|llllllll}
$c_2=\beta_2/\beta_0$ & $s_0$ & $s_1$ & $f_1^2$ & $s_2$ & $f_2^2$  & $\Lambda$ [GeV] & $M_0$ & $\A_1(0)$
\\ 
\hline
-5.73 & 25.01 & 18.220 & 0.3091 & 0.7082 & 0.6312 & 0.231 & 1.15 & 1.00 
\\
-4.76 & 23.06 & 16.837 & 0.2713 & 0.8077 & 0.5409  & 0.260 & 1.25 & 0.776
\\
-2.10 & 17.09 & 12.523 & 0.1815 & 0.7796 & 0.3462 & 0.363 & 1.50 & 0.544
\end{tabular}
\end{ruledtabular}
\end{table}

However, the world average of the coupling parameter \cite{PDG2010} has
some uncertainty:
$a^{({\overline {\rm MS}})}(M_Z^2) = (0.1184 \pm 0.0007)/\pi$,
corresponding to 
$a^{({\overline {\rm MS}})}(m_{\tau}^2)_{n_f=3} = (0.3183 \pm 0.0057)/\pi$, i.e., 
${\overline \Lambda}_{n_f=3} = 0.336 \pm 0.010$ GeV where 
${\overline \Lambda}_{n_f=3}$ is the usual $\MSbar$ scale (at $n_f=3$).
This implies that the Lambert scale varies,
$\Lambda \approx 0.260 \pm 0.008$ GeV
[$a(m_{\tau}^2,c_2) = (0.2905 \pm 0.0043)/\pi$], while all the (central) values
of the dimensionless parameters of the model ($c_2=-4.76, s_0=23.06$, etc.)
are unchanged. The predicted value of the
$V$+$A$ $\tau$-decay ratio then varies, $r_{\tau} = 0.203 \pm 0.006$, which
is still compatible with the experimental value
$0.203 \pm 0.004$ \cite{ALEPH2,DDHMZ}.

\begin{figure}[htb] 
\begin{minipage}[b]{.49\linewidth}
\centering{\epsfig{file=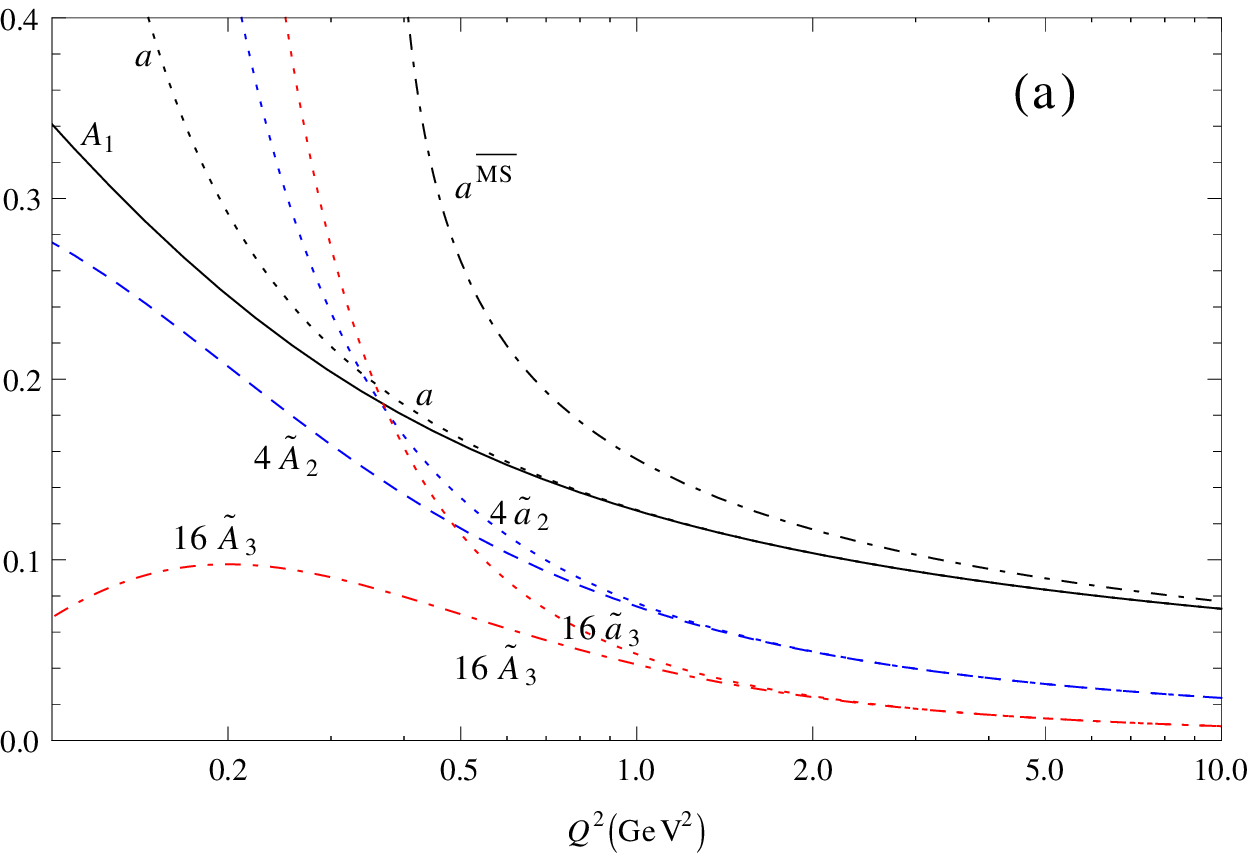,width=80mm,angle=0}}
\end{minipage}
\begin{minipage}[b]{.49\linewidth}
\centering{\epsfig{file=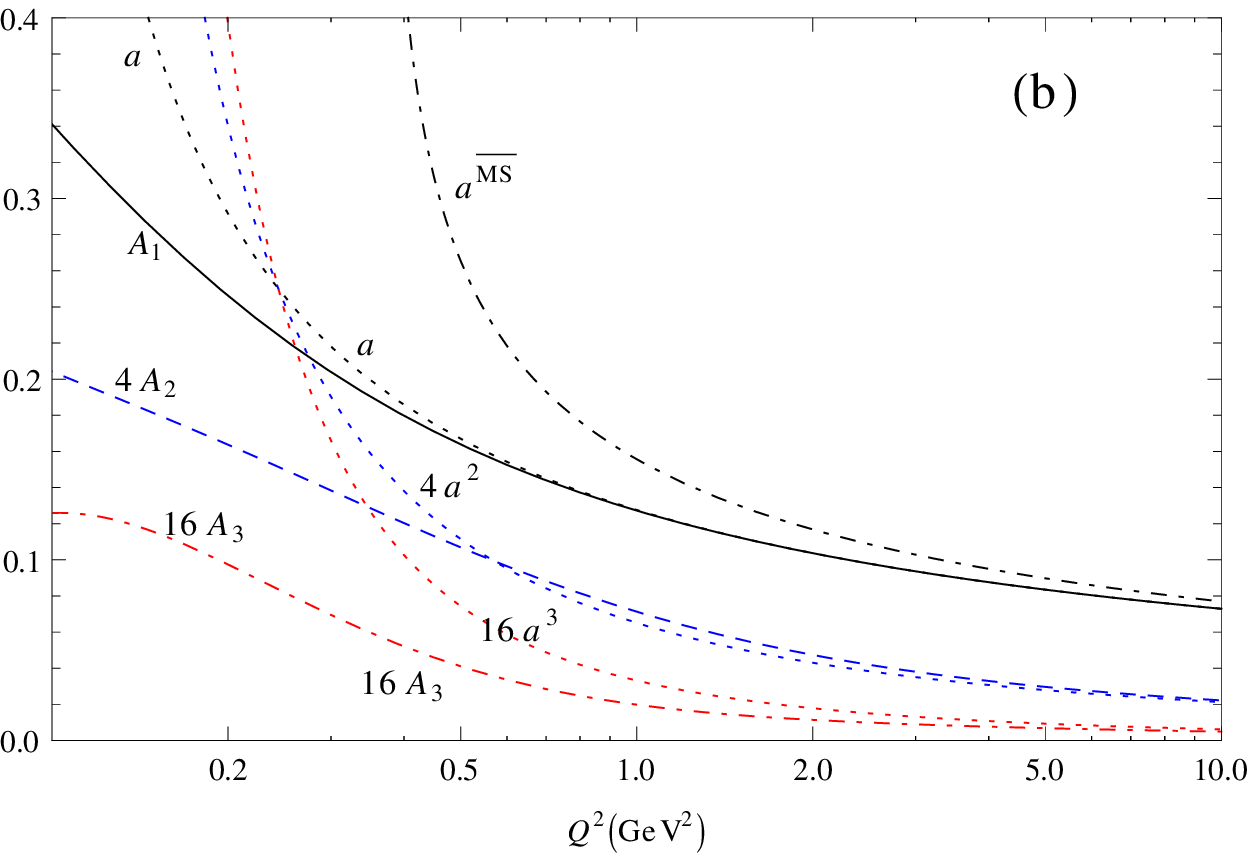,width=80mm,angle=0}}
\end{minipage}
\vspace{-0.4cm}
 \caption{(a) The couplings $\A_1(Q^2)$, $4 \times \tA_2(Q^2)$ and 
$4^2 \times \tA_3(Q^2)$ of 2-delta analytic QCD model with the central 
input values ($c_2=-4.76$; $s_0=23.06$), as a function of positive $Q^2$:
$0.1 \ {\rm GeV}^2 \leq Q^2 \leq 10.0 \ {\rm GeV}^2$; the
rescaling factors $4$ and $4^2$ are for better visibility; included are
the analogous pQCD couplings $a$, $4 \times \ta_2$ and 
$4^2 \times \ta_3$ in the
same (Lambert) scheme, and the $\MSbar$ coupling $a^{\MSbar}$.
(b) The couplings  $\A_1(Q^2)$, $4 \times \A_2(Q^2)$ and 
$4^2 \times \A_3(Q^2)$, where the couplings $\A_2$ and $\A_3$ are constructed
by Eqs.~(\ref{A2})-(\ref{A3A4}) with truncation at (including) $\tA_4$ term; included
are the (Lambert) pQCD analogs $a$, $4 a^2$ and $4^2 a^3$.}
\label{FigplLambQpos}
 \end{figure}

Having the analytic analog $(a(Q^2))_{\rm an} = \A_1(Q^2)$
in 2-delta QCD analytic model in Eq.~(\ref{A1Q2}),
the analytization of higher integer powers $(a^n)_{\rm an} = \A_n$
is performed according to the construction in
Refs.~\cite{GCCV} which is applicable to any analytic QCD model.
We briefly present it below. The basic idea is to introduce the 
logarithmic derivatives 
\be
{\ta}_{n+1}(Q^2)
\equiv \frac{(-1)^{n}}{\beta_0^{n} n!}
\frac{ \partial^n a(Q^2)}{\partial (\ln Q^2)^n} \ , 
\qquad (n=1,2,\ldots) \ .
\label{tan}
\ee
We note that ${\ta}_{n+1}(Q^2) = a(Q^2)^{n+1} + {\cal O}(a^{n+2})$
by RGE $\partial a(Q^2)/\partial \ln Q^2 = \beta(a(Q^2))$, where
beta function $\beta(a)$ has the pQCD expansion as given in
Eq.~(\ref{betapt}). 
Due to the linearity of analytization, it follows from
$(a(Q^2))_{\rm an} = \A_1(Q^2)$ the relation
$(\partial a(Q^2)/\partial \ln Q^2)_{\rm an} = \partial \A_1(Q^2)/\partial \ln Q^2$,
and thus in general
\be
\left( {\ta}_{n+1}(Q^2) \right)_{\rm an} = \tA_{n+1}(Q^2) \ ,
\label{tatA}
\ee
where 
\be
\tA_{n+1}(Q^2)
\equiv \frac{(-1)^n}{\beta_0^n n!}
\frac{ \partial^n \A_1(Q^2)}{\partial (\ln Q^2)^n} \ .
\qquad (n=1,2,\ldots) \ ,
\label{tAn}
\ee
and where $\A_1(Q^2)$ is given in our case in Eq.~(\ref{A1Q2}). An interesting aspect 
is that in virtually any analytic QCD model, including the present one,
we have a clear hierarchy $|\A_1(Q^2)| > |\tA_2(Q^2)| > |\tA_3(Q^2)| > \cdots$
not just for large $|Q^2|>\Lambda^2$, but for any $Q^2$, 
cf. curves in Fig.~\ref{FigplLambQpos}(a) (which are for the presented
model and at $Q^2>0$). 
This suggests the following approach to the
evaluation of any dimension-zero ($D=0$) contribution  ${\cal D}(Q^2)$
of a massless spacelike observable, such as Adler function, 
in analytic QCD. Let the perturbation
series (pt) of this quantity be
\bea
{\cal D}(Q^2)_{\rm pt} &=& a(\kappa Q^2)
\sum_{n=1}^{\infty} d_n(\kappa) \; a(\kappa Q^2)^{n+1} \ ,
\label{Dpt}
\eea    
where $\mu^2 = \kappa Q^2$ is a renormalization scale,
$\kappa \sim 1$ being a fixed chosen 
dimensionless renormalization scale parameter.\footnote{
For $\kappa$ dependence of $d_n(\kappa)$ coefficients,
see the next Section \ref{sec:resum}, Eqs.~(\ref{d1RScl})-(\ref{d3RScl}).}
Before evaluating ${\cal D}(Q^2)$ in analytic QCD, 
we reorganize the above series into the
corresponding ``modified'' perturbation series (mpt) in logarithmic
derivatives $\ta_{n+1}$, Eq.~(\ref{tan})
\bea
{\cal D}(Q^2)_{\rm mpt} &=& a(\kappa Q^2) + 
\sum_{n=1}^{\infty} {\td}_n(\kappa) \; {\ta}_{n+1}(\kappa Q^2) \ .
\label{Dmpt}
\eea
This leads, after applying the analytization (\ref{tatA})
term-by-term, to the (``modified'') analytic series (man)
\bea
{\cal D}(Q^2)_{\rm man} &=& \A_1(\kappa Q^2) + 
\sum_{n=1}^{\infty} {\td}_n(\kappa) \; \tA_{n+1}(\kappa Q^2) \ ,
\label{Dman}
\eea 
which is the basic expression for evaluation of ${\cal D}(Q^2)$ in
analytic QCD. 

Since the ``mpt'' series (\ref{Dmpt}) is just a reorganization of the
$\kappa$-independent ``pt'' series (\ref{Dpt}), the ``mpt'' series
${\cal D}(Q^2)_{\rm mpt}$ is also $\kappa$-independent. This then immediately
implies, in conjunction with the recurrence relation 
$\partial \ta_n(\kappa Q^2)/\partial \ln \kappa = - \beta_0 n \ta_{n+1}(\kappa Q^2)$
[this being a direct consequence of the definition (\ref{tan})],
the following set of differential relations between
$\td_n(\kappa)$:
\be
\frac{d}{d \ln \kappa} \td_n(\kappa) = n \beta_0 \td_{n-1} (\kappa) 
\qquad (n=1,2,\ldots) \ ,
\label{tdndiff}
\ee
where $d_0(\kappa)={\td}_0(\kappa) = 1$ by definition. 
Using these relations, it is straightforward to verify that the
(``modified'') analytic series ${\cal D}(Q^2)_{\rm man}$ of Eq.~(\ref{Dman})
is $\kappa$-independent.\footnote{
For $\kappa$ dependence of $\td_n(\kappa)$ coefficients,
obtained upon integrating the relations (\ref{tdndiff}),
see Eq.~(\ref{tdnmu}) in the next Section \ref{sec:resum}.}

The coefficients $\td_n(\kappa)$ are obtained from $d_k(\kappa)$'s
($k \leq n$) in the following way. We first relate powers
$a^{n+1}$ and the logarithmic derivatives $\ta_{n+1}$, at a given
scale $Q^2$ (or: $\mu^2 = \kappa Q^2$) using the RGE relations in
pQCD\footnote{
The first RGE is $\partial a(Q^2)/\partial \ln Q^2 = \beta(a(Q^2))$, where 
beta function $\beta(a)$ has pQCD expansion given in Eq.~(\ref{betapt}),
with $c_j \equiv \beta_j/\beta_0$. The other RGE equations are obtained by applying
further derivatives $\partial/\partial \ln Q^2$ to the first RGE.}
\bea
\ta_{2} &=& a^2 + c_1 a^3 + c_2 a^4 + \cdots \ ,
\label{ta2}
\\
\ta_{3} &=& a^3 + \frac{5}{2} c_1 a^4  + \cdots \ ,
\qquad
\ta_{4} = a^4 +   \cdots \ ,  
\qquad {\rm etc.} \ ,
\label{ta3ta4}
\eea 
and we invert them
\bea
a^2 & = & \ta_{2}
- c_1 \ta_{3} 
+ \left( \frac{5}{2} c_1^2 - c_2 \right) {\tilde a}_{4} + \cdots \ ,
\label{a2}
\\
a^3 & = & \ta_{3} - \frac{5}{2} c_1 \ta_{4} + \cdots \ ,
\qquad
 a^4  =  \ta_{4}  +  \cdots \ ,
\qquad {\rm etc.}
\label{a3a4}
\eea
Replacing the relations (\ref{a2})-(\ref{a3a4}) into the perturbation
expansion (\ref{Dpt}) for ${\cal D}(Q^2)$ we can read off the
tilde coefficients $\td_n(\kappa)$ of the reorganized (``modified'')
expansions (\ref{Dmpt})-(\ref{Dman})
\bea
\td_1(\kappa) & = & d_1(\kappa) \ , \qquad
\td_2(\kappa) = d_2(\kappa) - c_1 d_1(\kappa) \ ,
\label{td1td2}
\\
\td_3(\kappa) & = & d_3(\kappa) - \frac{5}{2} c_1 d_2(\kappa)
+ \left( \frac{5}{2} c_1^2 - c_2 \right) d_1(\kappa) \ ,
\qquad {\rm etc.}
\label{td3}
\eea
Applying analytization, Eqs.~(\ref{tatA})-(\ref{tAn}), in relations
(\ref{a2})-(\ref{a3a4}) term-by-term, we finally obtain the
analytic analogs of integer powers, $\A_n = (a^n)_{\rm an}$\footnote{
Expressions of $\tA_n$ in terms of $\A_k$'s ($k \geq n$), 
which can be used via inversion to obtain $\A_n$ in terms of
$\tA_k$'s, for integer $n$ and $k$, were given
in the context of MA model of \cite{ShS,MSS,Sh}
in Refs.~\cite{KuMa,Shirkov:2006nc,rev2}.}
\bea
\A_2 & \equiv & \left( a^2 \right)_{\rm an} 
= \tA_2 - c_1 \tA_3
+ \left( \frac{5}{2} c_1^2 - c_2 \right) \tA_{4} + \cdots \ ,
\label{A2}
\\
\A_3 & \equiv & \left( a^3 \right)_{\rm an} 
=  \tA_3 - \frac{5}{2} c_1 \tA_4 + \cdots \ ,
\quad
 \A_4 \equiv \left( a^4 \right)_{\rm an} 
=  \tA_{4}  +  \cdots \ ,
\qquad {\rm etc.}
\label{A3A4}
\eea
This means that the ``modified'' analytic series (\ref{Dman}) 
can be rewritten in the more usual form, in close analogy with the
original perturbation series (\ref{Dpt})
\bea
{\cal D}(Q^2)_{\rm an} &=& \A_1(\kappa Q^2) + 
\sum_{n=1}^{\infty} d_n(\kappa) \; \A_{n+1}(\kappa Q^2) \ .
\label{Dan}
\eea 
This series, being a reorganization of the 
$\kappa$-independent series ${\cal D}(Q^2)_{\rm man}$
of Eq.~(\ref{Dman}), is therefore also $\kappa$-independent.
Several couplings $\A_n$ in the present analytic QCD model, for 
low positive $Q^2$,
are presented in Fig.~\ref{FigplLambQpos}(b).

In practice, in the expansions of ${\cal D}(Q^2)$, 
Eqs.~(\ref{Dpt}) and (\ref{Dmpt}), we know exactly only a few first
coefficients $d_n(\kappa)$, 
up to (and including) $n=n_{\rm max}$, i.e., up to $a^{n_{\rm max}+1}$. 
This implies that in the relations (\ref{ta2})-(\ref{a3a4}) and
(\ref{A2})-(\ref{A3A4}) it is natural to perform truncations at
(and including) the term $\sim a^{n_{\rm max}+1}$ ($\sim \ta_{n_{\rm max}+1}
\sim \tA_{n_{\rm max}+1} \sim \A_{n_{\rm max}+1}$).
For example, in the case of Adler function, $n_{\rm max}=3$,
the perturbation series and the reorganized series are truncated
at $a^4$ ($\sim \ta_4$)
\bea
{\cal D}(Q^2;\kappa)_{\rm pt}^{[4]} &=& a(\kappa Q^2) + 
\sum_{n=1}^{3} d_n(\kappa) \; a(\kappa Q^2)^{n+1} \ ,
\label{Dtpt}
\\
{\cal D}(Q^2; \kappa)_{\rm mpt}^{[4]} &=& a(\kappa Q^2) + 
\sum_{n=1}^{3} {\td}_n(\kappa) \; {\ta}_{n+1}(\kappa Q^2) \ ,
\label{Dtmpt}
\eea
and the two analytic series (\ref{Dman}) and (\ref{Dan})
also become truncated, at $\A_4$ ($\sim \tA_4$)
\bea
{\cal D}(Q^2;\kappa)_{\rm man}^{[4]} &=& \A_1(\kappa Q^2) + 
\sum_{n=1}^{3} {\td}_n(\kappa) \; \tA_{n+1}(\kappa Q^2) \ ,
\label{Dtman}
\\
{\cal D}(Q^2;\kappa)_{\rm an}^{[4]} &=& \A_1(\kappa Q^2) + 
\sum_{n=1}^{3} d_n(\kappa) \; \A_{n+1}(\kappa Q^2) \ .
\label{Dtan}
\eea 
If the relations (\ref{ta2})-(\ref{a3a4}) and (\ref{A2})-(\ref{A3A4})
are then truncated naturally, i.e., at (and including)
$\sim a^4$ ($\sim \ta_4 \sim \tA_4 \sim \A_4$), it is straightforward
to check that the truncated series (\ref{Dtpt}) is identical with
(\ref{Dtmpt}), and (\ref{Dtman}) with (\ref{Dman}).

Due to the truncation, the above series are renormalization scale
($\kappa$) dependent. However, since the form of all the RGE relations of pQCD 
is maintained, by construction, also in analytic QCD under the
correspondence $\ta_n \mapsto \tA_n$ and $a^n \mapsto \A_n$, the
truncated analytic series (\ref{Dtman}) and (\ref{Dtan}) have
weak renormalization scale dependence of one order higher than the
last included term
\be
\frac{\partial {\cal D}(Q^2;\kappa)_{\rm man}^{[4]}}{\partial \ln \kappa}
\sim \tA_5(\kappa Q^2) \sim \A_5(\kappa Q^2) \sim
\frac{\partial {\cal D}(Q^2;\kappa)_{\rm an}^{[4]}}{\partial \ln \kappa}
\ ,
\label{RSdep1}
\ee
and this dependence is getting weaker when the number $N=n_{\rm max}+1$
of terms in the truncated series increases
\be
\frac{\partial {\cal D}(Q^2;\kappa)_{\rm man}^{[N]}}{\partial \ln \kappa}
\sim \tA_{N+1}(\kappa Q^2) \sim \A_{N+1}(\kappa Q^2) 
\sim \frac{\partial {\cal D}(Q^2;\kappa)_{\rm an}^{[N]}}{\partial \ln \kappa}
\ .
\label{RSdep2}
\ee
The last relation $\sim$ on the right side of Eq.~(\ref{RSdep2}) 
is valid as long as the couplings $\A_n$, appearing in 
${\cal D}(Q^2;\kappa)_{\rm an}^{[N]}$, are constructed via the linear
combinations Eqs.~(\ref{A2})-(\ref{A3A4}) with so many terms 
that the last term $\tA_M$ included there has $M \geq N$.
The derivative of the ``man'' truncated series on the
left-hand side of Eq.~(\ref{RSdep2}) is in fact only one term
\be
\frac{\partial {\cal D}(Q^2;\kappa)_{\rm man}^{[N]}}{\partial \ln \kappa}
= - \beta_0 N \td_{N-1}(\kappa) \tA_{N+1}(\kappa Q^2) \ ,
\label{RSdep3}
\ee
as can be explicitly checked by using the recursive relation\footnote{
This being a direct consequence of the definition (\ref{tAn}).}
$\partial \tA_{n+1}(\kappa Q^2)/\partial \ln \kappa = 
- \beta_0 (n+1) \tA_{n+2}(\kappa Q^2)$
and the differential relations (\ref{tdndiff}) between the
coefficients $\td_n(\kappa)$ which are
a consequence of $\kappa$ independence of the full ``man'' series
(\ref{Dman}).
The derivative of the ``an'' truncated series on the
right-hand side of Eq.~(\ref{RSdep2}) is in general
a finite linear combination of terms $\A_{R+1}(\kappa Q^2)$ with $R \geq N$,
the number of the terms of this combination depending on
the level of truncation made in the construction of $\A_n$'s
in the  relations (\ref{A2})-(\ref{A3A4}).
One of the benefits of using analytic QCD is that this
residual unphysical dependence 
$[\sim \tA_{N+1}(\kappa Q^2) \sim \A_{N+1}(\kappa Q^2)]$ is getting
weaker with increasing $N$ irrespective of the physical momentum scale
$Q^2$ (in stark contrast with truncated series in pQCD), because of
the aforementioned hierarchy  $|\A_1(Q^2)| > |\tA_2(Q^2)| > |\tA_3(Q^2)| > \cdots$
which is valid at any $Q^2$ (not just: $|Q^2| \gg \Lambda^2$).
The described method of analytization of integer powers $a^n$
was constructed in Refs.~\cite{GCCV}, and was extended in Ref.~\cite{GCAK}
to the case of terms with noninteger
powers $a^{\nu}$ and of terms of the form $a^{\nu} \ln^k a$.

If the analytization of higher powers $a^n$ were performed in the naive
nonlinear way, $(a^n)_{\rm an} = \A_1^n$, the renormalization scale dependence
of the truncated series would in practice not decrease with the
inclusion of more terms in the series,
but would in general even increase,
because in such a case the derivatives $\partial/\partial \ln \kappa$
of the truncated series include complicated nonperturbative 
contributions such as $1/(Q^2)^K$, cf.~Appendix C of Ref.~\cite{panQCD}
(second entry). The full (naive) analytic power series
${\cal D}(Q^2)_{\rm aps} = \A_1(\kappa Q^2) 
+ \sum_{n=1}^{\infty} d_n(\kappa) (\A_1(\kappa Q^2))^{n+1}$
is not $\kappa$-independent, basically because the powers $\A_1(\kappa Q^2)^n$
do not fulfill RGE relations analogous to those of $a(\kappa Q^2)^n$. 

The RGE of the coupling $\A_1(Q^2)$ has now, by construction, formally
the same form as the RGE of pQCD coupling $a(Q^2)$ where we
replace $a^n \mapsto \A_n$
\be
\frac{\partial \A_1(Q^2)}{\partial \ln Q^2} 
\left( \equiv - \beta_0 \tA_2(Q^2) \right) =
- \beta_0 \A_2(Q^2) - \beta_1 \A_3(Q^2) - \beta_2 \A_4(Q^2) - \ldots
\label{anRGE}
\ee 
In Sec.~\ref{sec:Borel} we will present the curves of the resulting
$\A_1(Q^2)$ and of the underlying Lambert pQCD coupling $a(Q^2)$ 
on the $Q^2$-contours in the complex plane which are needed
for the evaluation of the Borel sum rules there.

In the extraction of the experimental
value of the strangeless $V$+$A$
decay ratio $r_{\tau}(\triangle S=0, m_q=0)_{\rm exp} = 0.203 \pm 0.004$,
\cite{ALEPH2,DDHMZ}, the contributions of the  
higher dimension ($D=2,4,6,8$) chirality-violating terms
(i.e., nonzero quark mass effects) were subtracted.
These latter terms were estimated to be 
$\delta r_{\tau}(\triangle S=0, m_{u,d}\not=0) = (-5.8 \pm 1.4) \times 10^{-3}$
(cf.~also Appendix B of Ref.~\cite{panQCD}).
The chirality-nonviolating contributions were not subtracted
from  the mentioned $r_{\tau}(\triangle S=0, m_q=0)_{\rm exp} = 0.203 \pm 0.004$.
Among the chirality-nonviolating 
$D \geq 2$ contributions, the only possibly nonnegligible one 
(cf.~Ref.~\cite{Ioffe})
is the $D=4$ contribution from the gluon condensate,  
$( \delta r_{\tau} )_{\langle a GG \rangle} = 
(11/4) \alpha_s^2(m_{\tau}^2) \langle a G G \rangle/m_{\tau}^4$.
However, in our evaluation of $r_{\tau}(\triangle S=0, m_q=0)$ 
in the analytic QCD we assumed that the gluon condensate contribution is
negligible, i.e., $r_{\tau}(\triangle S=0, m_q=0)$ was evaluated as the
$D=0$ contribution only, and was required to achieve the value $0.203$.
Later in this article, we will deduce that the gluon condensate
in analytic QCD has similar values as in pQCD, i.e.,
$\langle a GG \rangle \approx 0.005 \ {\rm GeV}^4$, which then gives us
the contribution to $r_{\tau}$
\be
\left( \delta r_{\tau}^{(V+A)} \right)_{\langle a GG \rangle }
\approx \frac{11 \pi^2}{4} \tA_2(m_{\tau}^2) \frac{1}{m_{\tau}^4}
\langle a G G \rangle \approx 1.4 \times 10^{-4} \ ,
\label{drtauGG}
\ee
where we replaced $\alpha_s^2(m_{\tau}^2)/\pi^2 \equiv a(m_{\tau}^2)^2$
($\approx \ta_2(m_{\tau}^2)$) by $\tA_2(m_{\tau}^2) \approx 0.01$,
cf.~Fig.~\ref{FigplLambQpos}(a). Thus we can conclude a posteriori
that our neglecting of the higher dimension chirality-nonviolating
terms in the OPE of $r_{\tau}$ was justified in our evaluation of
$r_{\tau}$, where the latter evaluation contributed significantly to the
fixing of parameters of the (2-delta) analytic QCD model.

\section{Pad\'e-related resummation of Adler function in analytic QCD}
\label{sec:resum}

In this Section we summarize an evaluation method for massless spacelike
physical quantities, which we apply to the
evaluation of the dimension $D=0$ contribution of
Adler function, ${\cal D}(Q^2; D=0) \equiv {\cal D}(Q^2)$.
This method was developed some time ago for pQCD
evaluations \cite{BGApQCD1,BGApQCD2} and is a generalization of the
diagonal Pad\'e (dPA) resummation method in pQCD \cite{GardiPA}. 
We will first apply dPA to the mentioned ($D=0$) Adler
function ${\cal D}(Q^2)$. 

The perturbation series of this quantity
is known up to the fourth term, Eq.~(\ref{Dtpt}),
where $\mu^2= \kappa Q^2$ is the (squared) spacelike renormalization scale 
($\kappa \sim 1$), and the truncated series has a residual 
$\mu^2$-dependence due to truncation.
Coefficients ${\overline d}_n(1) \equiv d_j(1,\MSbar)$ ($n=1,2,3$),
in $\MSbar$ scheme and at the renormalization scale $\mu^2=Q^2$,
were obtained in Refs.~\cite{d1,d2,d3}, respectively
\bea
{\overline d}_1(1) & = & \frac{1}{12} + 0.691772 \; \beta_0 \ ,
\label{d1Adl}
\\
{\overline d}_2(1) &=& -27.849 + 8.22612 \; \beta_0 + 3.10345 \; \beta_0^2
\ ,
\label{d2Adl}
\\
{\overline d}_3(1) &=& 
32.727 - 115.199 \; \beta_0 + 49.5237 \; \beta_0^2 + 2.18004 \; \beta_0^3 \ .
\label{d3Adl}
\eea
The light-by-light contributions were not included in these
coefficients, since they are zero when the number of effective quark 
flavors is $n_f=3$ (then: $\beta_0=9/4$). 
The value $n_f=3$ is used in the evaluation of 
${\cal D}(Q^2)$ because the relevant energies in the analysis of the
next Section are $|Q^2| \alt m_{\tau}^2$
($m_{\tau}^2 \approx 3.2 \ {\rm GeV}^2 < (2 m_c)^2 \approx 6.5 \ {\rm GeV}^2$).

We are interested in evaluation of Adler function
in 2-delta analytic QCD model of Sec.~\ref{sec:model}
and in the corresponding pQCD with the same renormalization scheme.
Therefore, we have to transform first this expansion from the
$\MSbar$ renormalization scheme to the new (Lambert) scheme of
2-delta analytic QCD model: $c_2=-4.76; c_3=c_2^2/c_1;$ etc., 
cf.~second line of Table \ref{t1}, and 
Eqs.~(\ref{aptexact})-(\ref{betapt}) which define the
corresponding pQCD and the renormalization scheme. The
scheme invariance of the perturbation expansion (\ref{Dpt})
then implies that the coefficients in the new (Lambert) scheme are
\bea
d_1(1) & = & {\overline d}_1(1) \ , \qquad 
d_2(1) = {\overline d}_2(1) - (c_2 - {\overline c}_2 ) \ ,
\nonumber\\
d_3(1) & = & {\overline d}_3(1) - 
2 {\overline d}_1(1) (c_2 - {\overline c}_2 )
- \frac{1}{2} (c_3 - {\overline c}_3 ) \ , 
\label{RSchch}
\eea
where the bars denote the values in $\MSbar$ scheme.
The new expansion coefficients $d_j(\mu^2/Q^2)$, in
Lambert scheme and at the renormalization scale 
$\mu^2 = \kappa Q^2$, are then
\bea
d_1(\kappa) &=& d_1(1) + \beta_0 \ln(\kappa) \ ,
\label{d1RScl}
\\
d_2(\kappa) & = & d_2(1) + \sum_{k=1}^2 
\left(
\begin{array}{c}
2 \\
k
\end{array}
\right)
\beta_0^k  \ln^k ( \kappa ) d_{2-k}(1) + 
\beta_1 \ln ( \kappa ) \ ,
\label{d2RScl}
\\
d_3(\kappa) & = & d_3(1) + \sum_{k=1}^3 
\left(
\begin{array}{c}
3 \\
k
\end{array}
\right)
\beta_0^k \ln^k ( \kappa ) d_{3-k}(1) + 
\beta_1 \left[ 2 \ln ( \kappa )  d_1(1)
+ \frac{5}{2} \beta_0  \ln^2 ( \kappa ) \right] 
+ \beta_2 \ln ( \kappa ) \ ,
\label{d3RScl}
\eea
where these relations were obtained from the renormalization scale
invariance of the perturbation expansion (\ref{Dpt}).
The resulting truncated perturbation expansion 
${\cal D}(Q^2;\kappa)_{\rm pt}^{[4]}$, Eq.~(\ref{Dtpt}), is then used
as the basis for the evaluation of the $D=0$ Adler function 
${\cal D}(Q^2)$ in 2-delta analytic QCD model. Due to truncation,
it has (unphysical) renormalization scale dependence,
$\partial {\cal D}(Q^2;\kappa)_{\rm pt}^{[4]}/\partial \ln \kappa \sim a^5$,
and the same is true for the corresponding analytic truncated series
${\cal D}(Q^2;\kappa)_{\rm an}^{[4]}$, Eqs.~(\ref{Dtan}), i.e.,
$\partial {\cal D}(Q^2;\kappa)_{\rm an}^{[4]}/\partial \ln \kappa \sim \A_5$,
cf.~Eq.~(\ref{RSdep1}).

The dPA-resummed result can be  written in two equivalent ways\footnote{
We recall that in the case of a four-term power series (\ref{Dtpt}), 
the diagonal Pad\'e is ${\rm [2/2]}(a)$, which is by definition
the ratio of two quadratic polynomials 
in $a \equiv a(\mu^2)$ such that ${\cal D} -{\rm [2/2]}(a) \sim a^5$.}
\bea
[2/2]_{{\cal D}}(a(\mu^2)) &=& \frac{x (1 + E_1 x)}{1 + F_1 x + F_2 x^2}
{\bigg |}_{x=a(\mu^2)} 
\label{dPA1}
\\
&=& \left( \alpha_1 \frac{x}{1 + u_1 x} +
\alpha_2 \frac{x}{1 + u_2 x} \right) {\bigg |}_{x=a(\mu^2)} \ ,
\label{dPA2}
\eea
where $\alpha_1+\alpha_2=1$.
In Ref.~\cite{GardiPA} it was shown that this approximant is
independent of the renormalization scale $\mu^2 = \kappa Q^2$ 
(i.e., independent of $\kappa$) used in the original
truncated series (\ref{Dtpt}) if the RGE-running is at the one-loop level.
Building on this idea, in Refs.~\cite{BGApQCD1} this approach was
extended so that the $\mu^2$-independence of the 
resummed result was exact.\footnote{Since the physical quantity ${\cal D}(Q^2)$
is $\mu^2$-independent, this extended resummation, having the same
property, is expected to approximate
better the (unknown) full expression for ${\cal D}(Q^2)$.}
For this, the truncated perturbation series ${\cal D}(Q^2;\kappa)_{\rm pt}^{[4]}$,
Eq.~(\ref{Dtpt}), in powers of $a(\mu^2)$, was first reorganized
into the truncated series ${\cal D}(Q^2)_{\rm mpt}^{[4]}$, 
in logarithmic derivatives $\ta_{n+1}$ defined in Eq.~(\ref{tan}).
The factor in front of this definition was chosen so that 
${\ta}_1 \equiv a$ and  $\ta_{n+1} = a^{n+1} + {\cal O}(a^{n+2})$ for $n \geq 1$. 
We recall that at one-loop level $\ta_{n+1} = a^{n+1}$ 
(in general: $\ta_{n+1} \not= a^{n+1}$). 

The reorganized truncated 
(modified) perturbation series ${\cal D}(Q^2;\kappa)_{\rm mpt}^{[4]}$
is given in Eq.~(\ref{Dtmpt}),
where the new coefficients ${\td}_j(\kappa)$ are related to the original
coefficients $d_j(\kappa)$ in Eqs.~(\ref{td1td2})-(\ref{td3}),
and $c_j \equiv \beta_j/\beta_0$ are the coefficients 
of the beta function, Eq.~(\ref{RGE}).
The coefficients ${\td}_j(\kappa)$ can be regarded as ``the one-loop parts'' of
the original coefficients $d_j(\kappa)$, because they have the simple
one-loop type of renormalization scale dependence 
(involving only the $\beta_0$ coefficient). 
Namely, upon integrating directly 
the differential relations (\ref{tdndiff}),
which were obtained on the basis of $\kappa$ independence
of the full series ${\cal D}(Q^2)_{\rm mpt}$ of Eq.~(\ref{Dmpt}), we obtain
the explicit form of $\kappa$ dependence of $\td_n(\kappa)$
\be
{\td}_n(\kappa) = {\td}_n(1) + \sum_{k=1}^n 
\left(
\begin{array}{c}
n \\
k
\end{array}
\right)
\ \beta_0^k \ \ln^k ( \kappa ) {\td}_{n-k}(1) \ ,
\label{tdnmu}
\ee
where we recall that $\kappa$ is the dimensionless renormalization scale
parameter ($\kappa = \mu^2/Q^2$), and $d_0=\td_0=1$. The procedure for
the construction of the generalization of dPA method consists now in
the following. In the modified truncated series ${\cal D}(Q^2;\kappa)_{\rm mpt}^{[4]}$,
Eq.~(\ref{Dtmpt}), we replace, in the
one-loop sense, the logarithmic derivatives by the powers
$\ta_n(\mu^2) \mapsto a_{1 \ell}(\mu^2)^n$, and obtain a truncated power series
of a new quantity ${\widetilde {\cal D}}(Q^2)$
\be
{\widetilde {\cal D}}(Q^2;\kappa)_{\rm pt}^{[4]} \equiv 
a_{1\ell}(\kappa Q^2) + 
\sum_{n=1}^{3} {\td}_n(\kappa) \; a_{1\ell}(\kappa Q^2)^{n+1} \ ,
\label{tDpt}
\ee 
where $ a_{1 \ell}(\kappa Q^2)= a_{1 \ell}(\mu^2)$ 
is the coupling RGE-evolved from $a(Q^2)$ to 
$a(\mu^2)$ by the one-loop beta function $(- \beta_0 a^2)$
\be
a_{1\ell}(\kappa Q^2) = 
\frac{a(Q^2)}{1 + \beta_0 \ln(\kappa) \; a(Q^2)} \ .
\label{a1l}
\ee
The full (untruncated) series ${\widetilde {\cal D}}(Q^2) = a_{1\ell}(\kappa Q^2) + 
\sum_{n=1}^{\infty} {\td}_n(\kappa) \; a_{1\ell}(\kappa Q^2)^{n+1}$ is $\kappa$-independent,
as can be easily checked by using the 
differential relations (\ref{tdndiff}).
Now we apply to the truncated power series of ${\widetilde {\cal D}}(Q^2)_{\rm pt}$ 
the [2/2] Pad\'e approximant, as was performed in Eqs.~(\ref{dPA1})-(\ref{dPA2})
in the case of ${\cal D}(Q^2)$
\bea
[2/2]_{\widetilde {\cal D}}(a_{1\ell}(\kappa Q^2)) &=& 
\frac{x (1 + {\widetilde E}_1 x)}
{1 + {\widetilde F}_1 x + {\widetilde F}_2 x^2}
{\bigg |}_{x=a_{1\ell}(\kappa Q^2)} 
\label{dPA1tD}
\\
& = &
 \left( \tal_1 \frac{x}{1 + \tu_1(\kappa) x} +
\tal_2 \frac{x}{1 + \tu_2(\kappa) x} \right) 
{\bigg |}_{x=a_{1 \ell}(\kappa Q^2)} 
\label{dPA2tD}
\\
& = & \tal_1  \; a_{1 \ell}(\kappa_1 Q^2) + \tal_2  \; a_{1 \ell}(\kappa_2 Q^2) \ .
\label{dPA3tD}
\eea
Here, $\tal_1+\tal_2=1$. Going from Eq.~(\ref{dPA2tD}) to (\ref{dPA3tD}), we
used the relation (\ref{a1l}) and denoted the new scales as
\be
\tQ_j^2 = \kappa_j Q^2 \ ; \quad \kappa_j = \kappa 
\exp \left( \frac{\tu_j(\kappa)}{\beta_0} \right) \ ,
\qquad (j=1,2) \ .
\label{tQs}
\ee
The resulting approximant for the original truncated power series
(\ref{Dtpt}) is then obtained by simply replacing the one-loop
pQCD coupling $a_{1 \ell}$ in Eq.~(\ref{dPA3tD}) 
by the full pQCD coupling $a$ ($=\alpha_s/\pi$)
\be
{\cal G}^{[2/2]}_{{\cal D}}(Q^2) =  
\tal_1 \; a(\kappa_1 Q^2) +\tal_2 \; a(\kappa_2 Q^2) \ ,
\label{dBG22}
 \ee 
This method of construction can be applied in a completely analogous way
when the number $N$ of known perturbation terms in the series of ${\cal D}(Q^2)_{\rm pt}$
is any even number $N=2 M$ ($N=2, 4, 6, \ldots$),\footnote{
When $N=2$, the method reduces to the (two-loop) 
effective charge method of Refs.~\cite{ECH}.}
leading to the approximant
\be
{\cal G}^{[M/M]}_{{\cal D}}(Q^2) =  
\sum_{j=1}^M \tal_j \; a(\kappa_j Q^2) \ ,
\label{dBG}
\ee
where $\tal_1 + \ldots + \tal_M = 1$.
In Ref.~\cite{BGApQCD1} it was proven that the result is exactly independent
of the original renormalization scale $\mu^2 = \kappa Q^2$. 
In the proof in Ref.~\cite{BGApQCD1} it was demonstrated that
each weight coefficient $\tal_j$ and each
scale coefficient $\kappa_j$ is separately independent of the 
renormalization scale parameter $\kappa$; for a somewhat
less formal and more intuitive proof, see Appendix \ref{app} here.
The $\kappa$-independent coefficients $\tal_j$ and $\kappa_j \equiv \tQ_j^2/Q^2$
are also $Q^2$-independent since they are dimensionless.
In Ref.~\cite{BGApQCD1} it was also proven that the approximant fulfills the
basic approximation requirement required of any resummation approximant
\be
{\cal D}(Q^2) - {\cal G}^{[M/M]}_{{\cal D}}(Q^2) = {\cal O}(\ta_{2 M+1})
= {\cal O}(a^{2 M+1})
\ .
\label{dBGappr}
\ee   
The approximant (\ref{dBG22}), although theoretically attractive, turned
out not to work well within pQCD. The reason for this is that one of the
two scales in Eq.~(\ref{dBG22}), e.g.~$\tQ_1^2$, is lower than $Q^2$ and often
brings the pQCD coupling $a(\tQ_1^2)$ close to the Landau singularities,
making thus the evaluation of low-momentum physical quantities
${\cal D}(Q^2)$ in pQCD unreliable. For example, in $\MSbar$ scheme,
Adler function (with $n_f=3$) gives $\kappa_1 \ll 1$ (cf.~also footnote
\ref{bMSnot} in Sec.~\ref{sec:Borel}).

However, in Ref.~\cite{BGA} this method was revived and applied in
analytic QCD frameworks, where no such problems of Landau singularities appear.
It was demonstrated in Ref.~\cite{BGA} that the approximant (\ref{dBG22})
should be applied with the same weights and the same scales as in pQCD 
also in analytic QCD
where the analytic truncated power series of the physical quantity has
the form (\ref{Dtan}) analogous to Eq.~(\ref{Dtpt})
and the modified truncated analytic series (\ref{Dtman})
has the form analogous
to Eq.~(\ref{Dtmpt}).\footnote{
 The couplings $\tA_{j+1}(\mu^2)$ are obtained from $\A_1(\mu^2)$ in complete 
analogy with Eq.~(\ref{tan}), according to Eq.~(\ref{tAn}).
The construction of $\A_n(\mu^2)$'s, as a linear combination of these quantities,
is given in Eqs.~(\ref{A2})-(\ref{A3A4}), and
was explained in detail in Refs.~\cite{GCCV,GCAK}.}
The resummed result is completely analogous to Eq.~(\ref{dBG22})
\be
{\cal G}^{[2/2]}_{{\cal D}}(Q^2;{\rm an}) =  
\tal_1 \; \A_1(\kappa_1 Q^2)+ \tal_2 \; \A_1(\kappa_2 Q^2)  \ ,
\label{dBGan22}
 \ee
with $\tal_j$ and $\kappa_j$ obtained by construction 
in Eqs.~(\ref{dPA2tD}) and (\ref{tQs}) and thus $\mu^2$-
and $Q^2$-independent. The applicability of the approximant (\ref{dBGan22})
is based on the fact, proven in Ref.~\cite{BGA}, that it also
fulfills (the analytic analog of) the basic approximation requirement, i.e.,
\be
{\cal D}(Q^2) - {\cal G}^{[2/2]}_{{\cal D}}(Q^2;{\rm an.}) 
\sim \tA_5(Q^2) \sim  \A_5(Q^2) \ .
\label{dBGanappr}
\ee   
For the analogously constructed ${\cal G}^{[M/M]}_{{\cal D}}(Q^2;{\rm an.})$, 
the above difference
becomes $\sim \tA_{2 M+1}$ ($\sim \A_{2 M+1}$), cf.~Ref.~\cite{BGA}.

We recall that in pQCD we regard ${\cal D} = {\cal D}_{\rm pt} ={\cal D}_{\rm mpt}$,
and in analytic QCD ${\cal D} = {\cal D}_{\rm an} ={\cal D}_{\rm man}$, where
these series quantities are written in Eqs.~(\ref{Dpt})-(\ref{Dmpt}) and
in Eqs.~(\ref{Dan}) and (\ref{Dman}), respectively.

In the renormalization scheme of 2-delta analytic QCD model
(Lambert scheme central choice: $c_2=-4.76$, $c_j=c_2^{j-1}/c_1^{j-2}$ for $j \geq 3$),
where the three coefficients $d_j(\kappa)$ at general renormalization scale
parameter $\kappa$ ($\equiv \mu^2/Q^2$) are obtained via 
Eqs.~(\ref{d1Adl})-(\ref{d3RScl}), the described formalism gives us
the following values of the scale coefficients $\kappa_j$ and
weights $\tal_j$:
\bea
\kappa_1 \left( \equiv \tQ_1^2/Q^2 \right) &=& 0.1689 \ , \quad \tal_1=0.6418 \ ,
\nonumber\\
\kappa_2 \left( \equiv \tQ_2^2/Q^2 \right) &=& 3.1656 \ , \quad \tal_2=0.3582 \ .
\label{tQtal}
\eea
These quantities are each exactly independent of the choice of the
original renormalization scale parameter $\kappa$ ($\equiv \mu^2/Q^2$) 
in the original expansion coefficients (\ref{d1RScl})-(\ref{d3RScl}), 
as proven in Ref.~\cite{BGApQCD1} and in Appendix \ref{app}, and as
can be checked also numerically by starting with
the construction of these quantities from $d_j(\kappa^{\prime})$'s 
at a different $\kappa^{\prime}$.

The massless Adler function ${\cal D}(Q^2) \equiv {\cal D}(Q^2;D=0)$, 
which is a logarithmic
derivative of the leading-dimension part of the polarization operator 
(correlator) of hadronic currents, will play a central role 
in the next Section in the analysis of the Borel sum rules 
involving invariant-mass spectra of the $\tau$ lepton 
decay, applied within 2-delta analytic QCD model described in the
previous Section. We will thus apply the method of
resummation described here, Eq.~(\ref{dBGan22}), for the evaluation of the Adler function
at complex $Q^2$ ($|Q^2| \sim  m_{\tau}^2$).

\section{Analysis of tau decay data with Borel sum rules}
\label{sec:Borel}

The idea of sum rules in $\tau$ decay physics
could be summarized as an application of the identity
\be
\int_0^{\sigma_0} d \sigma f(-\sigma) \omega_{\rm exp}(\sigma)  =
-i \pi  \oint_{|Q^2|=\sigma_0} d Q^2 f(Q^2) \Pi_{\rm th}(Q^2)  \ ,
\label{sr1}
\ee
where the contour integration on the right-hand side is in the
counterclockwise direction,
$f(Q^2)$ is an analytic function in the $Q^2$ complex plane,
and $\omega(\sigma)$ is the spectral function of the polarization function
$\Pi(Q^2)$ of hadronic currents
\be
\omega(\sigma) \equiv 2 \pi \; {\rm Im} \ \Pi(Q^2=-\sigma - i \epsilon) \ .
\label{om1}
\ee
The identity (\ref{sr1}) is obtained by applying the Cauchy theorem
to the function $f(Q^2) \Pi(Q^2)$ and taking into account
the analytic properties of the physical polarization function $\Pi(Q^2)$ 
as required by the general principles
of quantum field theories. We recall that in 
pQCD-evaluated (pQCD+OPE) 
polarization function $\Pi(Q^2)_{\rm th}$ these analyticity properties 
are in general not respected, because of the Landau singularities of the
pQCD coupling $a(Q^2)$. This means that in pQCD, when we replace on the
left-hand side of Eq.~(\ref{sr1}) $\omega_{\rm exp}(\sigma) \mapsto
\omega_{\rm th}(\sigma)$, the identity in general ceases being valid.
In analytic QCD no such conceptual problems appear, the
theoretically evaluated $\Pi(Q^2)_{\rm th}$ automatically respects the analyticity
properties on which the sum rule (\ref{sr1}) is based.

In this work we are interested in the nonstrange $V$+$A$ channel of
$\tau$ decays. As a consequence, the polarization function is
a sum of functions
\be
\Pi(Q^2) = \Pi^{(1)}_{V}(Q^2) +  \Pi^{(1)}_{A}(Q^2) + \Pi^{(0)}_{(A)}(Q^2) \ .
\label{Pi1}
\ee 
These functions appear in the polarization operators $\Pi^J_{\mu\nu}(q)$
which are correlators of the (nonstrange) charged hadronic currents
\bea
\Pi^J_{\mu\nu}(q) & = & i \int  d^4 x \; e^{i q \cdot x} 
\langle T J_{\mu}(x) J_{\nu}(0)^{\dagger} \rangle
\nonumber\\
& = & (q_{\mu} q_{\nu} - g_{\mu \nu} q^2) \Pi_J^{(1)}(Q^2)
+ q_{\mu} q_{\nu} \Pi_J^{(0)}(Q^2) \ ,
\label{PiJ1}
\\
{\rm where:} \ J&=&V,A; \quad J_{\mu} = {\overline u} \gamma_{\mu} d \;\; 
(J=V) \ ,
\quad  J_{\mu} = {\overline u} \gamma_{\mu} \gamma_5 d \;\; (J=A) \ .
\label{PiJ2}
\eea
For more details on these points, we refer to 
Refs.~\cite{Geshkenbein,Ioffe}.
On the left-hand side of the rum rule (\ref{sr1}) the experimental
spectral function $\omega_{\rm exp}(\sigma)$ is used, obtained from
the measured $\tau$-decay invariant-mass spectra for 
$0 < \sigma < m_{\tau}^2$.\footnote{
The data published by ALEPH Collaboration are in Refs.~\cite{ALEPH1,ALEPH2}.
The experimental bands represented by
the left-hand side of Eq.~(\ref{sr1}), for $f(Q^2)=\exp(Q^2/M^2)/M^2$, are
taken here from Figs.~4 and 5(a),(b)
of Ref.~\cite{Geshkenbein}, which in turn were obtained from
the ALEPH Collaboration data of 1998, Ref.~\cite{ALEPH1}.
\label{expbands}} 
On the right-hand side of Eq.~(\ref{sr1}), the theoretically
evaluated polarization function $\Pi(Q^2)$ appears, which can be evaluated with
the OPE
\be
\Pi(Q^2)_{\rm th} =  - \frac{1}{2 \pi^2} \ln(Q^2/\mu^2) + 
\Pi(Q^2;D\!=\!0)
+ \sum_{n \geq 2} \frac{ \langle O_{2n} \rangle}{(Q^2)^n} \left(
1 + {\cal C}_n a(Q^2) \right) \ .
\label{OPE1}
\ee
The $D=2$ operator term (i.e., $n=1$ term) comes from the
nonzero values of the current masses of $u$ and $d$ quarks, it is negligible
and is neglected here.
For the evaluation of the contour integral on the
right-hand side of Eq.~(\ref{sr1}), it is convenient 
for us to perform integration by parts, resulting in
\be
\int_0^{\sigma_0} d \sigma f(-\sigma) \omega_{\rm exp}(\sigma) =
- \frac{i}{2 \pi}  \int_{\phi=-\pi}^{\pi} \frac{d Q^2}{Q^2}
{\cal D}_{\rm Adl}(Q^2) \left[{\cal F}(Q^2) - {\cal F}(-s_0) \right] {\big |}_{Q^2 = s_0 \exp(i \phi)} \ ,
\label{sr2}
\ee
where ${\cal D}_{\rm Adl}(Q^2)$ is full Adler function, i.e., 
\bea
{\cal D}_{\rm Adl}(Q^2) &\equiv&  - 2 \pi^2 \frac{d \Pi(Q^2)}{d \ln Q^2} 
\label{Adl1}
\\
& = & 1 + {\cal D}(Q^2;D\!=\!0) + 2 \pi^2 \sum_{n \geq 2}
 \frac{ n \langle O_{2n} \rangle}{(Q^2)^n} \left(
1 + {\cal C}_n a(Q^2) \right) \ ,
\label{OPE2}
\eea
and the function ${\cal F}$ is any function satisfying the relation
\be
\frac{d {\cal F}(Q^2)}{d Q^2} = f(Q^2) \ .
\label{Ff}
\ee
The dimension zero ($D=0$, or leading-twist) terms are related by 
\be
\left( {\cal D}(Q^2) \equiv \right) \; 
{\cal D}(Q^2;D\!=\!0) =  - 2 \pi^2 \frac{d \Pi(Q^2;D\!=\!0)}{d \ln Q^2} \ .
\label{Adl2}
\ee
The perturbation expansion of the $D=0$ massless and strangeless 
Adler function ${\cal D}(Q^2) \equiv {\cal D}(Q^2;D\!=\!0)$, cf.~Eq.~(\ref{Dtpt}),
is now known up to $\sim a^4$ order
with the coefficients ${\overline d}_j(1)$ ($j=1,2,3$) in $\MSbar$ scheme
and at renormalization scale $\mu^2=Q^2$ given 
in Eqs.~(\ref{d1Adl})-(\ref{d3Adl}) for the here relevant momentum regime
$|Q^2| \alt m_{\tau}^2$ (i.e., for $n_f=3$).
We transformed the expansion from $\MSbar$ 
renormalization scheme to the new (Lambert) scheme of
2-delta analytic QCD model 
($c_2=-4.76; c_3=c_2^2/c_1;$ etc., cf.~second line of Table \ref{t1})
according to the relations (\ref{RSchch}), and the resulting
coefficients $d_j(\kappa)$, at an arbitrary renormalization scale parameter
$\kappa \equiv \mu^2/Q^2$, are given in Eqs.~(\ref{d1RScl})-(\ref{d3RScl}). 
The resulting truncated analytic series, in 2-delta analytic QCD, 
is given in Eqs.~(\ref{Dtman})-(\ref{Dtan}).
The latter two expressions are identical because the construction of
the analytic analogs $(a^n)_{\rm an} \equiv \A_n$ was performed as a 
truncated linear combination
of the logarithmic derivatives $\tA_k$'s of $\A_1$ [cf.~Eq.~(\ref{tAn})]
according to Eqs.~(\ref{A2})-(\ref{A3A4}), with the last included 
term in those linear combinations being $\tA_4$.
On the other hand, the resummed expression was obtained in Eq.~(\ref{dBGan22}) 
in 2-delta analytic QCD and Eq.~(\ref{dBG22}) in pQCD, 
with the RGE-invariant values of the scale and weight coefficients
$\kappa_j$ and $\tal_j$ given in Eqs.~(\ref{tQtal}). We refer for more
details to Sections \ref{sec:model} and \ref{sec:resum}. 

The basic idea of the Borel sum rules is to choose, in the sum rule
relation (\ref{sr1}) [or: (\ref{sr2})], 
for the function $f$ an exponential function \cite{Geshkenbein,Ioffe}
\be
f(Q^2) = \frac{1}{M^2} \exp( Q^2/M^2) \ , \qquad
{\cal F}(Q^2) = \exp( Q^2/M^2) \ ,
\label{fFBorel}
\ee
where $M^2$ are, in principle, arbitrary complex scales. 
Other choices of function $f$, in the context of sum rules,
have also been used in the literature, 
cf.~Refs.~\cite{Ioffe:2000ns,varsumrules,O6sr1,O6sr2,O4sr1}.
The integrals in the sum rules (\ref{sr1}), (\ref{sr2}), with the choice
(\ref{fFBorel}), become  Borel transforms\footnote{
We choose in the sum rules (\ref{sr1}) and (\ref{sr2})  for the upper integration
bound the largest possible value, i.e., $\sigma_0 = m_{\tau}^2$. If $\sigma_0$ is 
significantly below $m_{\tau}^2$, it has been argued in the literature
that in such a case the duality-violating effects become important 
and have to be accounted for, cf.~Ref.~\cite{DV}.}
$B(M^2)$, and the sum rule acquires the form
\be
B_{\rm exp}(M^2) =  B_{\rm th}(M^2) \ ,
\label{sr3a}
\ee
where
\bea
B_{\rm exp}(M^2) &\equiv& \int_0^{m_{\tau}^2} 
\frac{d \sigma}{M^2} \; \exp( - \sigma/M^2) \omega_{\rm exp}(\sigma) \ ,
\label{sr3b}
\\
B_{\rm th}(M^2) &\equiv& B(M^2;D\!=\!0) + 2 \pi^2 \sum_{n \geq 2}
 \frac{ \langle O_{2n} \rangle}{ (n-1)! \; (M^2)^n} \ ,
\label{sr3c}
\eea
and the $D=0$ part is the following contour integral:
\bea
B(M^2;D\!=\!0) & = & \frac{1}{2 \pi}\int_{-\pi}^{\pi}
d \phi \; {\cal D}(Q^2\!=\!m_{\tau}^2 e^{i \phi};D\!=\!0) \left[ 
\exp \left( \frac{m_{\tau}^2 e^{i \phi}}{M^2} \right) -
\exp \left( - \frac{m_{\tau}^2}{M^2} \right) \right] \ .
\label{BD0}
\eea
The Borel transform suppresses the $D \geq 4$ terms by $(n-1)!$ 
factor ($n \equiv D/2$),\footnote{
We checked that the terms ${\cal C}_n a(Q^2) \langle O_{2 n} \rangle/(Q^2)^n$
in the OPE expansion (\ref{OPE1}) give negligible contributions
and were thus ingored in the Borel sum rules (\ref{sr3a})-(\ref{sr3c}).}
and suppresses the high energy tail of $\omega_{\rm exp}(\sigma)$
where the experimental errors are larger. For the low-energy regime
$|Q^2| < 1 \ {\rm GeV}^2$, it does not provide suppression.
In Refs.~\cite{Geshkenbein,Ioffe} it was argued that the $D=0$ part of the
Borel sum rule can be reliably calculated within pQCD 
only for $|M|^2 > 0.8$-$1 \ {\rm GeV}^2$,
due to the (unphysical) Landau singularities of the pQCD coupling
$a(Q^2)$. In analytic QCD we do not have this problem.

\begin{figure}[htb] 
\begin{minipage}[b]{.49\linewidth}
\centering\includegraphics[width=85mm]{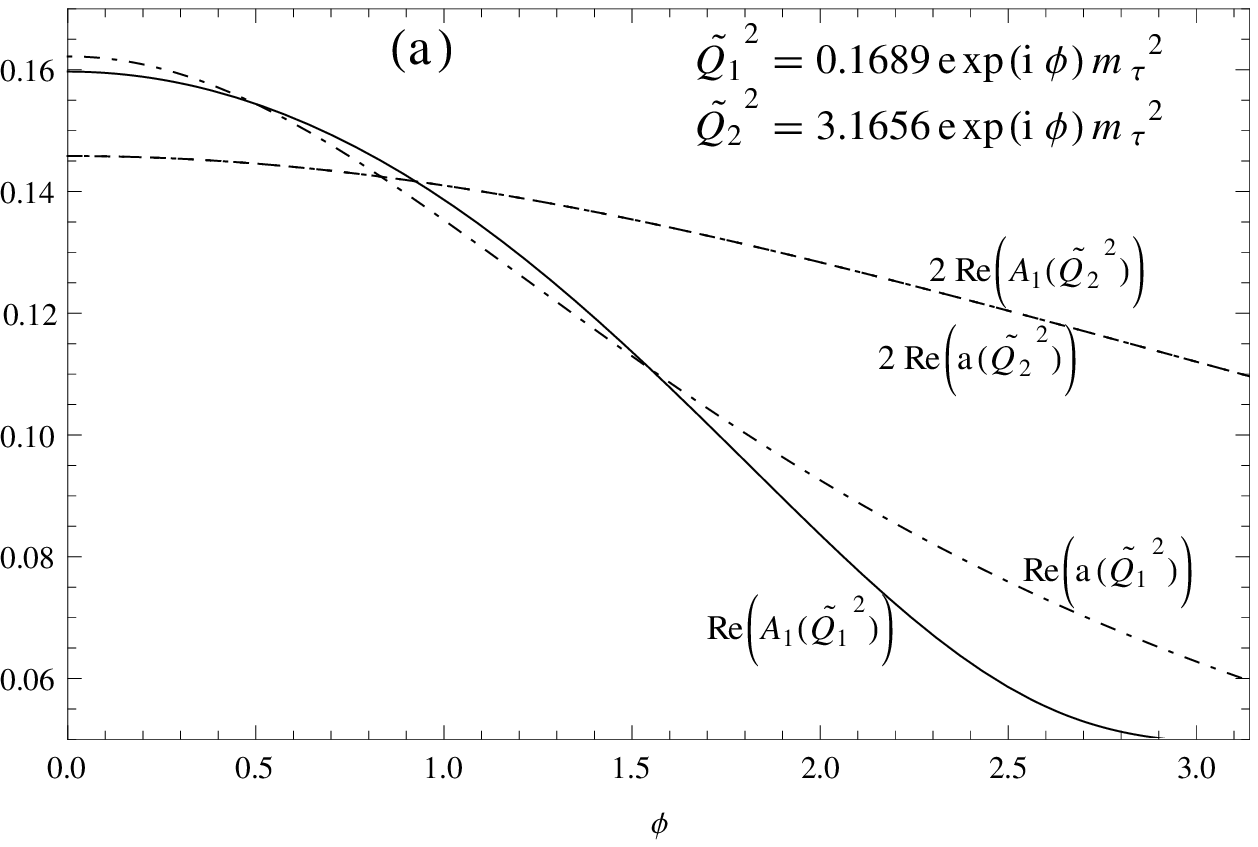}
\end{minipage}
\begin{minipage}[b]{.49\linewidth}
\centering\includegraphics[width=85mm]{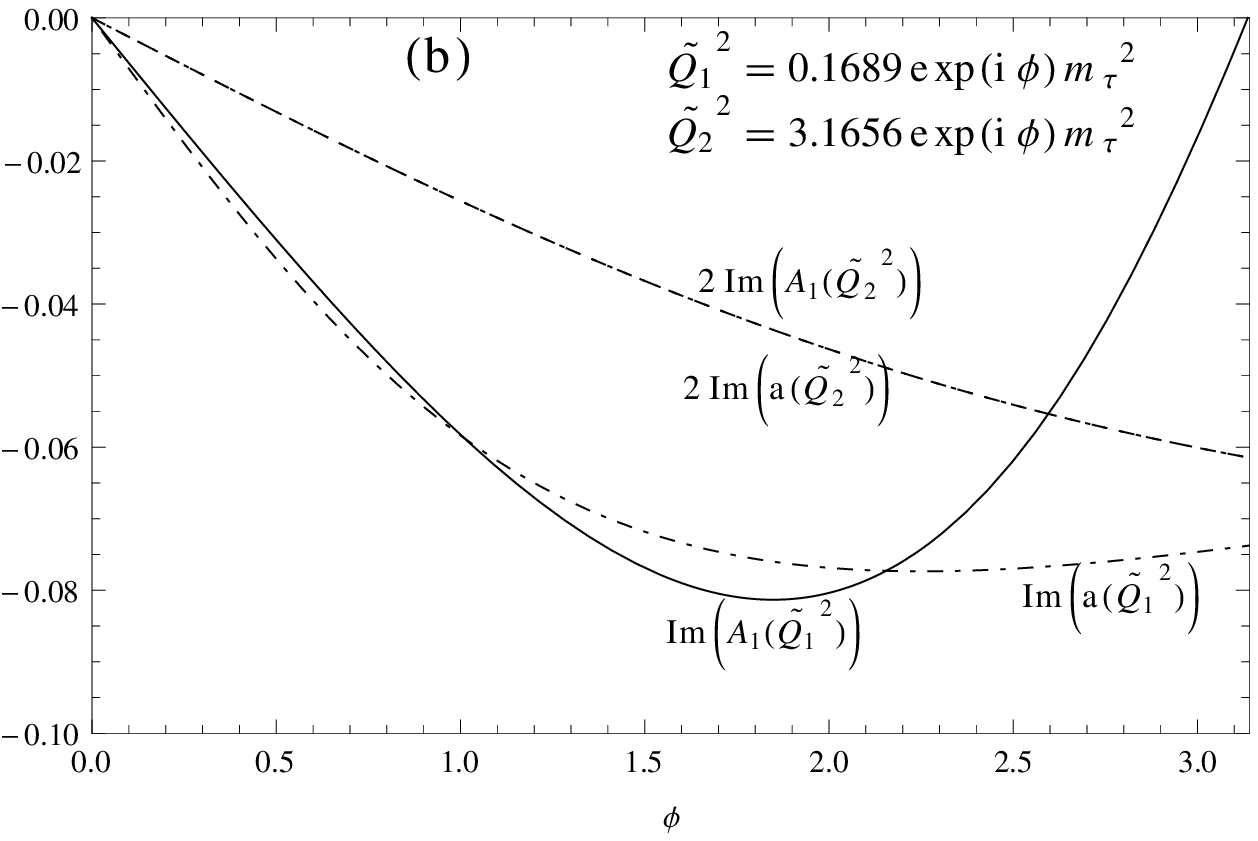}
\end{minipage}
\vspace{-0.4cm}
\caption{ (a) Real and (b) imaginary part of
the analytic coupling $\A_1(\tQ_j^2)$ and of the underlying pQCD
coupling $a(\tQ_j^2)$ ($j=1,2$) for the contour $Q^2=m_{\tau}^2 \exp(i \phi)$.
At the higher scales ($\tQ_2^2$), the two couplings are indistinguishable.
The scheme parameter value is $c_2=-4.76$ (Lambert scheme,
cf.~the second line of Table \ref{t1}).}
\label{FigplLamb}
 \end{figure}

Eq.~(\ref{BD0}) implies that we have to evaluate the Adler function
$D(Q^2;D\!=\!0)$ along the contour $Q^2 = m_{\tau}^2 \exp( i \phi)$. In particular,
if resummation is performed, 
Eqs.~(\ref{dBGan22}) and (\ref{dBG22}) imply that the
relevant coupling parameters in the Adler function
are $\A_1(\kappa_j m_{\tau}^2 \exp(i \phi))$ and $a(\kappa_j m_{\tau}^2 \exp(i \phi))$,
with $\kappa_1$ and $\kappa_2$ as given in Eq.~(\ref{tQtal}). 
The real and imaginary parts of these couplings are presented in 
Figs.~\ref{FigplLamb} (a),(b). 
We can see from these Figures that the
analytic coupling $\A_1$ and the underlying pQCD coupling $a$
differ from each other appreciably only at the
low scales, i.e., at $\tQ_1^2 \equiv 0.1689 \ m_{\tau}^2 \exp(i \phi)$.

\begin{figure}[htb] 
\begin{minipage}[b]{.49\linewidth}
\centering\includegraphics[width=85mm]{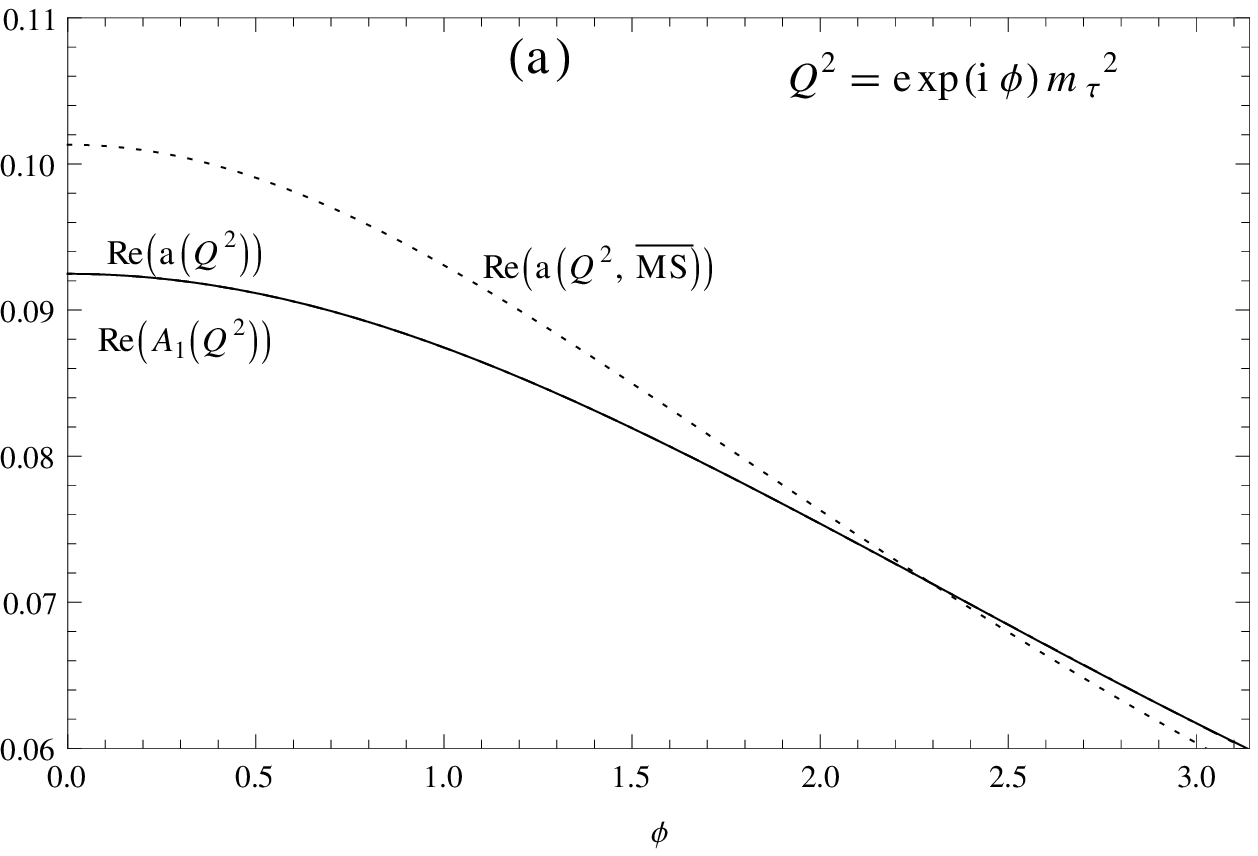}
\end{minipage}
\begin{minipage}[b]{.49\linewidth}
\centering\includegraphics[width=85mm]{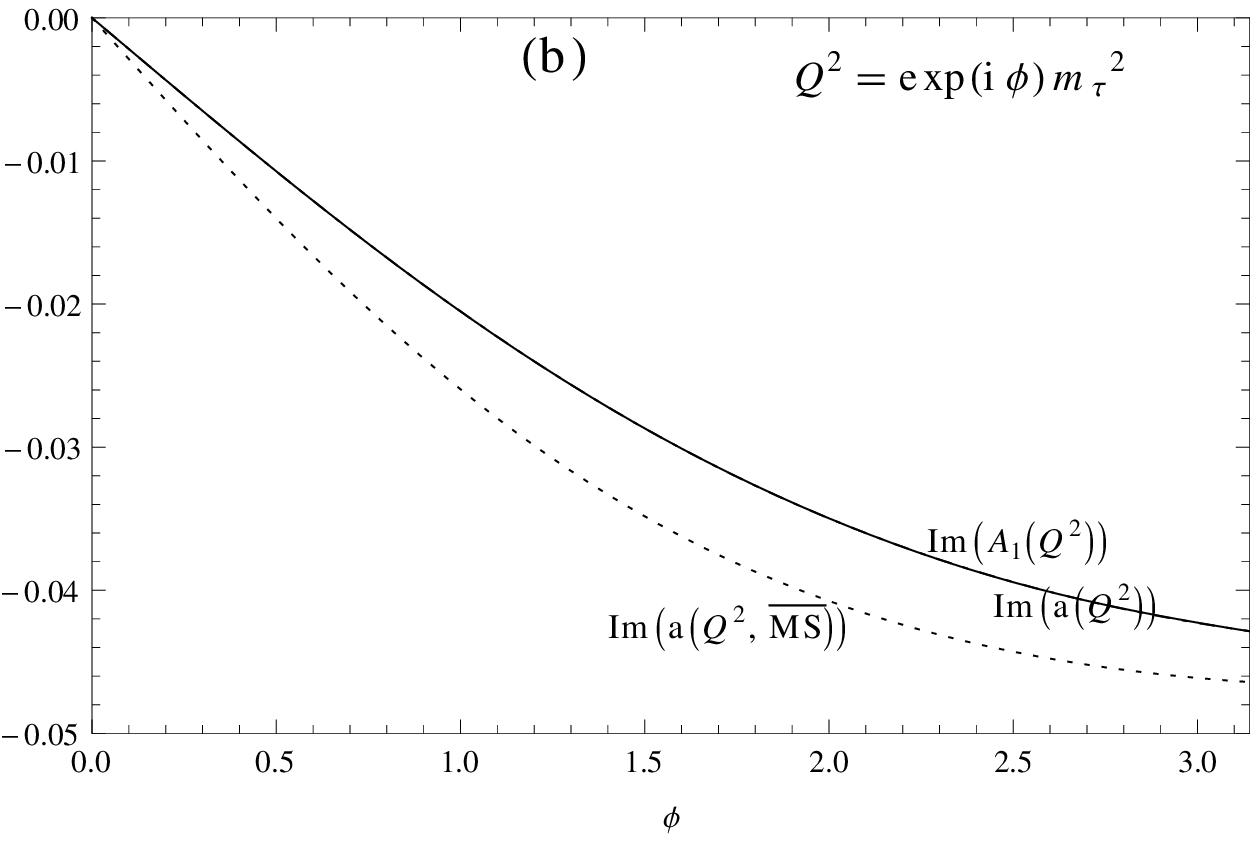}
\end{minipage}
\vspace{-0.4cm}
\caption{  (a) Real and (b) imaginary part of
the analytic coupling $\A_1(Q^2)$ and of the 
underlying pQCD coupling $a(Q^2)$ 
for the contour $Q^2=m_{\tau}^2 \exp(i \phi)$, with $c_2=-4.76$.
The two couplings are practically indistiguishable.
Included is also $\MSbar$ coupling ($c_2 \approx 4.47$, $c_3 \approx 20.99$) 
along the contour.}
\label{Figpls}
 \end{figure}

On the other hand, if not performing the resummation in analytic QCD, 
the truncated series (\ref{Dtman}) and thus the logarithmic derivatives
$\tA_{j+1}(\mu^2)$ have to be evaluated; and in the
underlying pQCD, the powers of $a(\mu^2)$ 
have to be evaluated, cf.~Eq.~(\ref{Dtpt}).
In all such cases, the otherwise arbitrary
renormalization scale $\mu^2=\kappa Q^2$ will be
chosen with $\kappa=1$, i.e., $Q^2 = m_{\tau}^2 \exp(i \phi)$. The real
and imaginary parts of the couplings
$\A_1(Q^2)$ and $a(Q^2)$, as well as of the $\MSbar$ coupling $a(Q^2;\MSbar)$,
as functions of the contour angle $\phi$, are presented in 
Figs.~\ref{Figpls} (a),(b). 
We can see that the analytic coupling $\A_1(Q^2)$
and the underlying pQCD coupling $a(Q^2)$ are almost
indistinguishable at such scales $|Q^2| = m_{\tau}^2$.
In fact, also the corresponding logarithmic derivatives $\ta_{j+1}(Q^2)$
and $\tA_{j+1}(Q^2)$ are very close to each other.\footnote{
Since we have in 2-delta analytic QCD the relation
$\A_1(Q^2) - a(Q^2) \sim (\Lambda^2/Q^2)^5$ (for $|Q^2| > \Lambda^2$),
repeated application of $\partial/\partial \ln Q^2$ to this relation
gives $\tA_{j+1}(Q^2) - \ta_{j+1}(Q^2) \sim (\Lambda^2/Q^2)^5$.}
This means that the two ``modified'' truncated
approaches (\ref{Dtmpt}) and (\ref{Dtman}), with
$\mu^2=Q^2$ ($=m_{\tau}^2 \exp(i \phi)$), give us results very close 
to each other. While in analytic QCD we will truncate the series
in terms of logarithmic derivatives, Eq.~(\ref{Dtman}),
we will, however, in pQCD apply the truncation to
the power series, Eq.~(\ref{Dtpt}) (with $\mu^2=Q^2$), instead.
Therefore, due to this somewhat different kind of truncation, 
the difference will be appreciable, but small nonetheless,
between the truncated analytic and the truncated pQCD approaches.

In order to separate or isolate terms of various dimensions in the 
Borel sum rule (\ref{sr3a})-(\ref{sr3c}), 
the Borel transform is evaluated along
fixed chosen rays in the complex $M^2$ plane \cite{Geshkenbein,Ioffe} 
\be
M^2 = |M|^2 \exp(i \psi) \ , \quad 
0.68 \ {\rm GeV}^2 < |M|^2 < 1.5 \ {\rm GeV}^2 \ .
\label{rays}
\ee
For example, when $\psi=\pi/6$ ($\psi=\pi/4$), 
the real part of the Borel transform contains
no $D=6$ ($D=4$) term\footnote{We include in the OPE sum 
(\ref{OPE1})-(\ref{OPE2})
only terms up to $n=3$, i.e., dimension $D \equiv 2 n = 6$.}
because ${\rm Re}(e^{i \pi/2})=0$
\bea
{\rm Re} B_{\rm exp}(|M|^2 e^{i \pi/6} ) &=& 
{\rm Re} B(|M|^2 e^{i \pi/6};D\!=\!0) + \pi^2 \frac{ \langle O_4 \rangle}{|M|^4}
\ ,
\label{pi6}
\\
{\rm Re} B_{\rm exp}(|M|^2 e^{i \pi/4} ) &=& 
{\rm Re} B(|M|^2 e^{i \pi/4};D\!=\!0) 
- \pi^2 \frac{ \langle O_6 \rangle}{\sqrt{2} |M|^6}
\ ,
\label{pi4}
\eea
The $D=4$ and $D=6$ operators can be expressed in terms of
condensates \cite{Shifman:1978bx}
\bea
\langle O_4^{(V+A)} \rangle &=& \frac{1}{6} \langle a G^{\alpha}_{\mu \nu}
G^{\alpha}_{\mu \nu} \rangle \ ,
\label{aGG1}
\\
\langle O_6^{(V+A)} \rangle  & \approx & 
\frac{128 \pi^2 }{81} a \langle {\overline q} q \rangle^2 \ ,
\label{aqq1}
\eea
where $a = \alpha_s/\pi$. The approximation (\ref{aqq1}) for
$\langle O_6^{(V+A)} \rangle$ is obtained after
factorization (vacuum saturation)
assumption of various 4-quark condensate contributions, and
is expected to be valid with not better than $20$-$30\%$ accuracy 
\cite{Shifman:1978bx,Geshkenbein}.
\begin{figure}[htb] 
\centering\includegraphics[width=140mm]{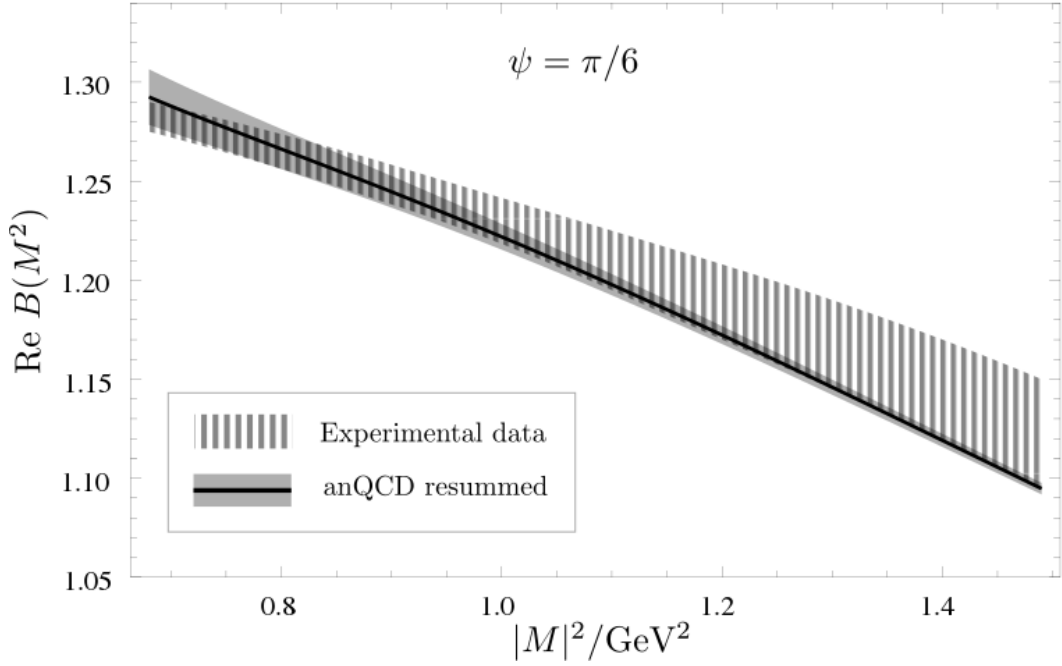}
\vspace{0.1cm}
 \caption{  Real part of the Borel transform, 
Eqs.~(\ref{sr3a})-(\ref{sr3c}) and (\ref{pi6}), 
for $M^2 = |M|^2 \exp(i \psi)$ with $\psi=\pi/6$.
The band with dark vertical stripes 
represents the experimental data (see footnote \ref{expbands}). 
The solid line is the best theoretical curve
of the resummed expression for $D=0$ Adler function ${\cal D}(Q^2)$ 
in 2-delta analytic QCD model (anQCD resummed),
obtained by the standard minimization with respect to the 
central experimental values. It corresponds  
to the central value of the gluon condensate $\langle a GG \rangle = 0.0055 \ {\rm GeV}^4$. 
The light grey band represents the variation of this curve
when the standard minimization is applied with respect to the values
of the upper and lower bounds of the experimental band, and gives
the variation of the gluon condensate $0.0055 \pm 0.0040 \ {\rm GeV}^4$. 
} 
\label{FigaGG}
 \end{figure}

The results of our analysis for the $\psi=\pi/6$ are given in Fig.~\ref{FigaGG}.
Using 2-delta analytic QCD model described in Sec.~\ref{sec:model}
and the resummation method of Sec.~\ref{sec:resum}
for the evaluation of the $D=0$ part, Eq.~(\ref{BD0}), we obtain for the
gluon condensate the values
\be
 \langle a G^{\alpha}_{\mu \nu} G^{\alpha}_{\mu \nu} \rangle =
(0.0055 \pm 0.0040_{\rm (exp)} \pm 0.0025_{\rm (oth)}) \ {\rm GeV}^4 = 
0.0055 \pm 0.0047 \ {\rm GeV}^4 \ .
\label{aGGres}
\ee
For the parameters of the analytic QCD model we used the central values of the
parameters as determined in Ref.~\cite{2danQCD}, i.e., 
the second line of Table \ref{t1} here. 
The central condensate value $0.0055 \ {\rm GeV}^4$ was obtained by
adjusting the gluon condensate value in such a way that the theoretical
curve (the solid line in Fig.~\ref{FigaGG}) gives the minimal deviation
from the central experimental values,\footnote{
At a given value of $|M|^2$, the central experimental value was considered to
be the arithmetic average of the upper and the lower bound value of the
experimental band at that $|M|^2$.}
in the standard minimization procedure,
using $40$ equidistant points over the depicted $|M|^2$-interval. 
The resulting deviation parameter was $\chi^2 = 6.40$. An interesting feature of
the resulting best theoretical curve (solid line) is that it remains within
the experimental band in almost the entire considered interval of $|M|^2$,
only slightly surpassing the upper bound of the experimental band
at the lowest values of $|M|^2$ ($|M|^2 \approx 0.68 \ {\rm GeV}^2$).

The uncertainty $\delta \langle a GG \rangle_{\rm exp} = \pm 0.0040 \ {\rm GeV}^4$
was obtained by applying the same minimization procedure to the
upper bounds and lower bounds of the experimental band, respectively.
The resulting two theoretical curves then define the upper and the
lower border of the light grey band in Fig.~\ref{FigaGG}. 
This band is thus the prediction
of the method (resummed analytic QCD), without the other uncertainties
included. 

The other uncertainty $\delta \langle a GG \rangle_{\rm oth} = \pm 0.0028 \ {\rm GeV}^4$
was obtained as the variation of the resummed analytic QCD prediction
when the  QCD coupling parameter value 
is varied in its world average interval \cite{PDG2010},
$\alpha_s^{(\MSbar)}(M_Z^2) = 0.1184 \mp 0.0007$; and when the
scheme parameter $c_2$ of the analytic model is varied
according to Table \ref{t1}, $c_2=-4.76 _{-0.97}^{+2.66}$.
At the end of Sec.~\ref{sec:model}, these two variations
in the model are discussed in more detail.\footnote{
$\alpha_s^{(\MSbar)}(M_Z^2) = 0.1184$ corresponds to: 
$a^{({\overline {\rm MS}})}(m_{\tau}^2)_{n_f=3} =0.3183/\pi$
(and to the standard $\MSbar$ scale: ${\overline \Lambda}_{n_f=3} = 0.336$ GeV);
and in Lambert scheme with $c_2=-4.76$ to $a(m_{\tau}^2) = 0.2905/\pi$
(and to Lambert scale $\Lambda = 0.260$ GeV).
On the other hand, varying the renormalization scale
$\mu^2 = \kappa Q^2 = \kappa m_{\tau}^2 \exp(i \phi)$ ($\leftrightarrow$ varying
$\kappa$) in the coefficients $d_j(\kappa)$ of the
expansion (\ref{Dtpt}) does not change at all the 
resummed results for the $D=0$ Adler function 
${\cal D}(Q^2)$ appearing 
in the contour integral (\ref{BD0}), as argued in Sec.~\ref{sec:resum}.}
The calculations give us for uncertainties of the value of 
the gluon condensate coming from these two effects:
$\delta \langle a GG \rangle_{\rm \alpha_s} =_{-0.0025}^{+0.0024} \ {\rm GeV}^4$,
and
$\delta \langle a GG \rangle_{\rm c_2} = _{+0.0004}^{+0.0001} \ {\rm GeV}^4$.
Adding in quadrature then gives us the ''other'' uncertainty  
$\delta \langle a GG \rangle_{\rm oth} \approx \pm 0.0025 \ {\rm GeV}^4$.

\begin{figure}[htb] 
\centering\includegraphics[width=140mm]{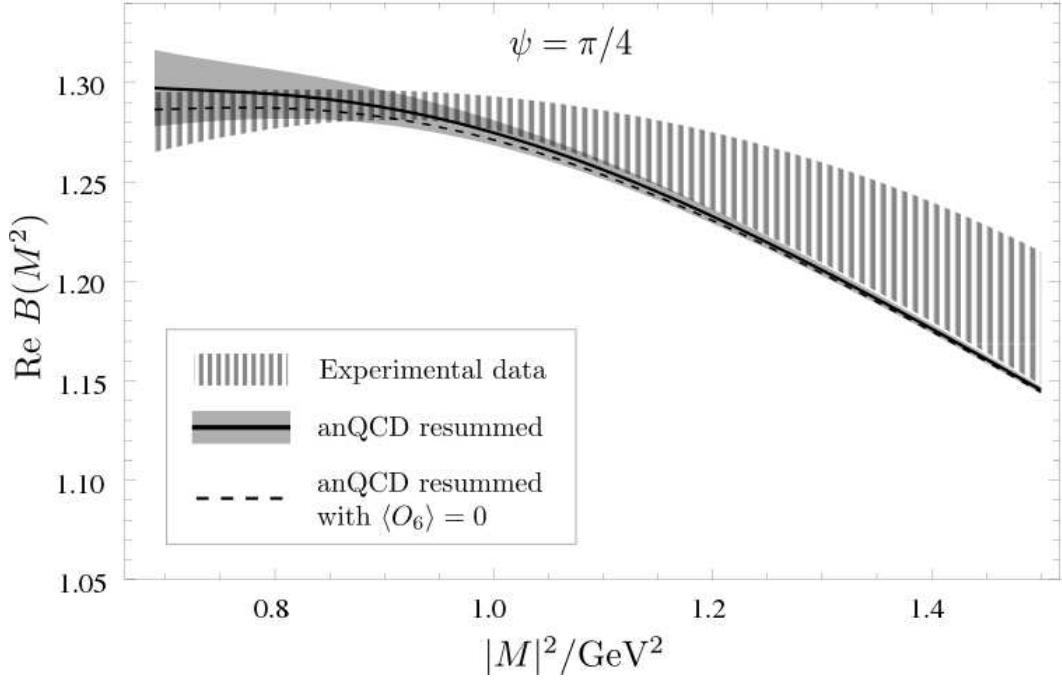}
\vspace{0.0cm}
 \caption{  Real part of the Borel transform, 
Eqs.~(\ref{sr3a})-(\ref{sr3c}) 
and (\ref{pi4}), for $M^2 = |M|^2 \exp(i \psi)$ with $\psi=\pi/4$. 
The band with dark vertical stripes represents the experimental data (see footnote \ref{expbands}). 
The best theoretical curve (solid line) is
obtained by standard minimization of the resummed analytic QCD results 
with respect to the central experimental values. 
The light grey band is obtained by applying the same minimization
to the upper and lower bounds of the experimental band.
The dashed curve is the prediction (of the resummed analytic QCD) when $\langle O_6^{(V+A)} \rangle=0$. 
See the text for details.}
\label{Figaqq}
 \end{figure}
If we adjust the gluon condensate values,
by the same standard minimization method (40 points)
with respect to the experimental values, 
in various evaluation approaches [anQCD resummed; anQCD (truncated); 
Lambert pQCD resummed; Lambert pQCD (truncated); $\MSbar$ (truncated)], 
we obtain the following values for the gluon condensates, 
and the corresponding deviation parameter $\chi^2$:
\bea
\langle a GG \rangle
&=&  (0.0055 \pm 0.0047) \ {\rm GeV}^4 \ , \; \chi^2=6.40 \ , \quad {\rm (anQCD \ resummed)} \ ,\
\nonumber\\
&=&  (0.0104 \pm 0.0058) \ {\rm GeV}^4 \ , \; \chi^2=9.60 \ , \quad {\rm (anQCD \ truncated)} \ ,\
\nonumber\\
&=& (0.0056 \pm 0.0049) \ {\rm GeV}^4 \ , \; \chi^2=7.42 \ , \quad  {\rm (Lambert \ resummed)} \ ,
\nonumber\\
&=& (0.0122 \pm 0.0046) \ {\rm GeV}^4 \ , \; \chi^2=9.83 \ , \quad  {\rm (Lambert \ truncated)} \ ,
\nonumber\\
&=& (0.0059 \pm 0.0049)  \ {\rm GeV}^4 \ , \; \chi^2=7.43 \ , \quad  (\MSbar \ {\rm truncated}) \ .
\label{aGGcentr}
\eea
The uncertainties given above were obtained by adding in
quadrature the experimental uncertainty 
$\delta \langle a GG \rangle_{\rm exp} = \pm 0.0040 \ {\rm GeV}^4$
and the uncertainty  from $\delta \alpha_s^{(\MSbar)}(M_Z^2) = \mp 0.0007$
[$\Rightarrow \ \delta \alpha_2^{(\MSbar)}(m_{\tau}^2)_{n_f=3}= \mp 0.0057$]
which is $\delta \langle a GG \rangle_{\rm \alpha_s} = \pm 0.0025$, $\pm 0.0025$, 
$\pm 0.0028$, $\pm 0.0023$, $\pm 0.0028 \ {\rm GeV}^4$,
for the respective five methods above. In the two analytic QCD 
approaches (anQCD resummed, anQCD truncated) the aforementioned uncertainty
coming from $c_2=-4.76 _{-0.97}^{+2.66}$ is  
$\delta \langle a GG \rangle_{\rm c_2}=_{+0.0004}^{+0.0001} \ {\rm GeV}^4$ and
$_{+0.0012}^{-0.0034} \ {\rm GeV}^4$, respectively, and was also added in quadrature.
We checked that the gluon condensate values change by less than $2 \%$ 
if the number of points in the standard minimization is decreased from $40$ to $28$;
the $\chi^2$ values decrease, but the values of ratios of $\chi^2$ between different
methods are maintained.

It is interesting that the evaluation with the pQCD approach in 
Lambert scheme (resummed) and in $\MSbar$ scheme (not resummed),\footnote{
\label{bMSnot}
We did not apply the resummation approach of Sec.~\ref{sec:resum} to
$D=0$ Adler function in pQCD in $\MSbar$ scheme. Namely, it turns out that  
$\MSbar$ is such a scheme which requires for the Adler function 
that the lower of the two
invariant scales be very low: $|\tQ_1| \ll 1 \ {\rm GeV}$
(while $|Q|=m_{\tau} \approx 1.78 \ {\rm GeV}$), leading to
severe problems with the Landau singularities. In the Lambert scheme,
on the other hand,
we obtain $|\tQ_1| \approx 0.73 \ {\rm GeV}$ and $|\tQ_2| \approx 3.16 \ {\rm GeV}$.}
give similar results for the gluon condensate as in 
anQCD resummed (and with only somewhat higher $\chi^2$).

However, the latter is not the case for the curves when 
$\psi=\pi/4$ [cf.~Eq.~(\ref{pi4})].
The results for the case $\psi=\pi/4$ are presented in Fig.~\ref{Figaqq}.
The central theoretical curve (solid line) 
represents the resummed analytic QCD result when the
condensate value $\langle O_6^{(V+A)} \rangle$ is obtained with the 
standard minimization (with 40 equidistant points),
in complete analogy with the $\psi=\pi/6$ case of Fig.~\ref{FigaGG}.
This central value is  $\langle O_6^{(V+A)} \rangle = -0.5 \times 10^{-3} \ {\rm GeV}^6$.
The resulting deviation parameter is $\chi^2 = 10.79$.
The best theoretical curve (solid line) only slightly surpasses the
upper bound of the experimental band at low $|M|^2 < 0.7 \ {\rm GeV}^2$,
remains within the experimental band
in the interval $0.7 \ {\rm GeV}^2 < |M|^2 < 1.0 \ {\rm GeV}^2$,
and is situated slightly below the lower bound of the experimental band
for $1.0 \ {\rm GeV}^2 < |M|^2$. It is interesting that the theoretical curve
with $\langle  O_6^{(V+A)} \rangle = 0$ (dashed line) is, in comparison,
not singificantly worse: it is situated slightly below the lower 
bound of the experimental band for $0.9 \ {\rm GeV}^2 < |M|^2$, and its
deviation parameter is $\chi^2 = 12.65$.

The uncertainty $\delta \langle  O_6^{(V+A)} \rangle_{\rm exp} = 
\pm 0.9 \times 10^{-3} \ {\rm GeV}^6$
is obtained by applying the same minimization procedure to the
upper bounds and lower bounds of the experimental band, respectively.
The resulting two theoretical curves then define the 
the light grey band in Fig.~\ref{Figaqq}, in analogy with
the light grey band of Fig.~\ref{FigaGG}.

The other uncertainty of $D=6$ condensate is estimated
as coming from the coupling parameter $\alpha_s$ 
and scheme parameter $c_2$ uncertainties, 
which are: $\delta \langle O_6^{(V+A)} \rangle_{\alpha_s} =
\pm 0.4 \times 10^{-3} \ {\rm GeV}^6$ and 
 $\delta \langle O_6^{(V+A)} \rangle_{c_2} =
\pm 0.4 \times 10^{-3} \ {\rm GeV}^6$. Adding in quadrature this
gives us $\delta \langle O_6^{(V+A)} \rangle_{\rm oth} = 
\pm 0.6 \times 10^{-3} \ {\rm GeV}^6$.
We thus obtain the following estimate for $D=6$ condensate of the $V$+$A$ channel:
\be
\langle O_6^{(V+A)} \rangle =
( -0.5 \pm 0.9_{\rm (exp)} \pm 0.6_{\rm (oth)} )
  \times 10^{-3} \ {\rm GeV}^6  \approx 
(-0.5 \pm 1.1) \times 10^{-3} \ {\rm GeV}^6 \ .
\label{O6res}
\ee
The standard minimization gave for the central value a negative number
close to zero ($-0.5 \times 10^{-3}  \ {\rm GeV}^6$). 
The result (\ref{O6res}) 
corresponds approximately to the factorized quark condensate values of
Eq.~(\ref{aqq1})
\be
a \langle {\overline q} q \rangle^2 \approx \frac{81}{128 \pi^2}
\langle O_6^{(V+A)} \rangle \approx (-3. \pm 7.) \times 10^{-5} \ {\rm GeV}^6 \ .
\label{aqqres2}
\ee
We note that the estimates (\ref{O6res})-(\ref{aqqres2}) are 
not incompatible with nonnegative values. Further, they are not incompatible
even with the following positive values of 
$a \langle  {\overline q} q \rangle^2$ extracted from the 
$V$-$A$ sum rules of Ref.~\cite{Ioffe}\footnote{
$V$-$A$ sum rules are more adequate to determine the quark condensate
$a  \langle  {\overline q} q \rangle^2$, because in such a case the factorization
approximation, in contrast to the $V$+$A$ case Eq.~(\ref{aqq1}), 
does not involve subtractions of large terms.}:
\bea
\langle O_6^{(V-A)} \rangle &=& - (4.4 \pm 0.6) \times 10^{-3} \ {\rm GeV}^6
\\
\Rightarrow \; 
a \langle  {\overline q} q \rangle^2 \left[ 
\approx - \frac{9}{64 \pi^2 \times 1.33} \langle O_6^{(V-A)} \rangle \right]
& \approx & (4.7 \pm 0.6) \times 10^{-5} \ {\rm GeV}^6 \ .
\label{VmA}
\eea
It is interesting that the methods other then the resummed analytic QCD 
approach give us for the central value of $D=6$ condensate 
significantly more negative values.
Specifically, if we adjust in these methods the $D=6$ condensate value 
by the standard minimization (with 40 equidistant points)
to the experimental values,
we obtain the following values of the condensate, and of the
$\chi^2$ fitting parameter:
\bea
\langle O_6^{(V+A)} \rangle 
&=&  (-0.5 \pm 1.1) \times 10^{-3} \ {\rm GeV}^6 \ , \; \chi^2=10.79 \ , \quad {\rm (anQCD \ resummed)} \ ,\
\nonumber\\
&=&  (-1.9 \pm 1.3) \times 10^{-3} \ {\rm GeV}^6 \ , \; \chi^2=15.75 \ , \quad {\rm (anQCD \ truncated)} \ ,\
\nonumber\\
&=&  (-1.8 \pm 0.9) \times 10^{-3} \ {\rm GeV}^6 \ , \; \chi^2=12.23 \ , \quad {\rm (Lambert \ resummed)} \ ,
\nonumber\\
&=&  (-2.3 \pm 0.9) \times 10^{-3} \ {\rm GeV}^6 \ , \; \chi^2=16.15 \ , \quad {\rm (Lambert \ truncated)} \ ,
\nonumber\\
&=&  (-1.8\pm 0.9)  \times 10^{-3} \ {\rm GeV}^6 \ , \; \chi^2=12.38 \ , \quad (\MSbar \ {\rm truncated}) \ .
\label{O6centr}
\eea
The uncertainties above include, in quadrature, the
experimental uncertainty 
$\delta \langle  O_6 \rangle_{\rm exp} = \pm 0.9 \times 10^{-3} \ {\rm GeV}^6$,
and the uncertainty from $\delta \alpha_s^{(\MSbar)}(M_Z^2) = \pm 0.0007$,
which is $\delta \langle  O_6 \rangle_{\alpha_s} = \pm 0.4 \times 10^{-3} \ {\rm GeV}^6$
for the two analytic QCD methods, and $ \pm 0.3 \times 10^{-3} \ {\rm GeV}^6$
for the three pQCD methods.
In the two analytic QCD approaches (anQCD resummed, anQCD truncated) 
the uncertainty coming from $c_2=-4.76 _{-0.97}^{+2.66}$ is 
$\delta \langle O_6 \rangle_{\rm c_2}=( _{-0.2}^{+0.4}) \times 10^{-3} \ {\rm GeV}^6$ and
$( _{-0.2}^{+0.8}) \times 10^{-3} \ {\rm GeV}^6$, respectively, and 
was also included in quadrature.
The results of all these methods, with the exception of
our central method (anQCD resummed),
are incompatible with nonnegative values of $\langle O_6^{(V+A)} \rangle$.
This agrees also with the analyses of $\tau$-decay data
in Refs.~\cite{O6sr1,O6sr2}, where
finite energy sum rules were applied (in pQCD+OPE approach)
in $\MSbar$ scheme. The results (\ref{O6centr}) thus suggest that the 
factorization assumption leading to the relation (\ref{aqq1})
fails to predict correctly even the sign of the condensate,
i.e., that it fails much more severely than by $20$-$30\%$
mentioned in Ref.~\cite{Geshkenbein}. On the other hand,
the properly resummed analytic QCD (+ OPE) approach gives us the
result (\ref{O6res}) and (\ref{aqqres2}), suggesting that the relation
(\ref{aqq1}) does not necessarily fail so severely. In fact, 
the resummed analytic QCD gives for the choice  $\langle O_6^{(V+A)} \rangle =0$
the value $\chi^2=12.65$, which is quite similar to the values
$\chi^2=12.38, 12.23$ obtained by the minimization in
$\MSbar$ pQCD and Lambert resummed pQCD, respectively (for 
the central value  
$\langle O_6^{(V+A)} \rangle \approx -1.8 \times 10^{-3} \ {\rm GeV}^6$).

When we increased the number of (equidistant in $|M|^2$) points in the
standard minimization procedure from 40 to 60, the values
of $\langle O_6^{(V+A)} \rangle$ in Eqs.~(\ref{O6centr}) changed very little,
and did not affect the digits displayed there; and the corresponding
values of $\chi^2$ increased proportionally, numerically by about $49$-$50 \%$.


Finally, when $\psi=0$, the Borel transform is real and includes,
in principle, both condensate contributions ($D=4, 6$).
\begin{figure}[htb] 
\centering\includegraphics[width=140mm]{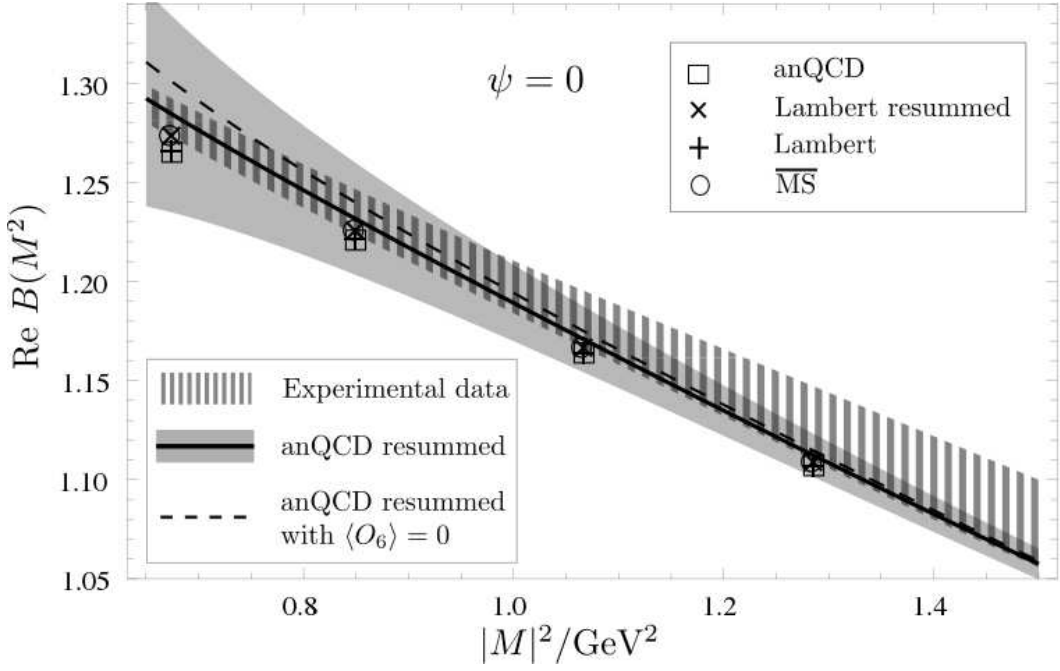}
\vspace{0.0cm}
 \caption{ 
The Borel transform for real $M^2 = |M|^2$ ($\psi=0$). 
The band with dark vertical stripes represents the experimental data (see footnote \ref{expbands}). 
The light grey band represents the variation
of the resummed 2-delta analytic QCD model prediction 
(anQCD resummed) when $D=4$ and $D=6$ condensates vary
in quadrature around their central values $0.0055 \ {\rm GeV}^4$
and $-0.5 \times 10^{-3} \ {\rm GeV}^6$, Eqs.~(\ref{aGGres}) and (\ref{O6res}).
The central solid curve is for the central values of both condensates.
The dashed curve is for the choice $\langle O_6^{(V+A)} \rangle =0$
(and $\langle a GG \rangle = 0.0055 \ {\rm GeV}^4$).
See the text for details. The results of other methods are given as points, for their corresponding central values of $D=4$ and $D=6$ condensates.}
\label{Figpsi0}
 \end{figure}
It is presented in Fig.~\ref{Figpsi0}. The result of the analytic QCD
(with resummed $D=0$ Adler function), for the central values of the
condensates, Eqs.~(\ref{aGGres}) and (\ref{O6res}), is represented 
by the central solid curve which is within
the experimental band in the entire interval of $M^2$.
The effects of variation of the values of the two condensates, 
Eqs.~(\ref{aGGres}) and (\ref{O6res}), 
are also presented in Fig.~\ref{Figpsi0} by the light grey band;
the total variations of the two condensates,
$\delta \langle O_{2n} \rangle$ ($n=2,3$), 
are added in quadrature [cf.~Eq.~(\ref{sr3c})]:
\be
B(M^2)_{\rm light \ grey} = B(M^2;D\!=\!0) + 2 \pi^2 \sum_{n=2,3}
 \frac{ \langle O_{2n} \rangle }{ (n-1)! \; (M^2)^n} 
\pm  2 \pi^2 \left[ \left( \frac{ \delta \langle O_4 \rangle }{M^4}
\right)^2 + \left( \frac{ \delta \langle O_6 \rangle }{2! M^6}
\right)^2 \right]^{1/2} \ .
\label{Blightgrey}
\ee
The dashed curve is the result of the choice $\langle O_6^{(V+A)} \rangle =0$
(and $\langle a GG \rangle = 0.0055 \ {\rm GeV}^4$).
The results of other methods, with their corresponding central values
of the condensates, Eqs.~(\ref{aGGcentr}) and (\ref{O6centr}), are included
as points in Fig.~\ref{Figpsi0}. These results are close to the lower
bound of the experimental band for all $M^2$. The resulting $\chi^2$ values
(with 40 points), 
with respect to the central experimental results (with $\psi=0$) are:
$\chi^2= 4.44$ (anQCD resummed); $\chi^2= 13.44$ (anQCD);  $\chi^2= 7.34$ (Lambert resummed);  $\chi^2= 13.84$ (Lambert);  $\chi^2= 7.58$ ($\MSbar$).
It is interesting that the dashed line (anQCD resummed with 
$\langle O_6^{(V+A)} \rangle =0$ and $\langle a GG \rangle = 0.0055 \ {\rm GeV}^4$) 
gives $\chi^2 = 7.02$, which is even slightly less than $\MSbar$ 
(with its own best central values:
$\langle O_6^{(V+A)} \rangle = - 1.8 \times 10^{-3} \ {\rm GeV}^6$ and 
$\langle a GG \rangle = 0.0059 \ {\rm GeV}^4$).
 
\section{Conclusions}
\label{sec:concl}

In this work we reanalyzed the Borel sum rules along the
rays in the $M^2$-complex plane, for the strangeless
$V$+$A$ $\tau$-decay invariant-mass spectra of ALEPH Collaboration
of 1998.
The analysis was performed here with the analytic
QCD model developed in Ref.~\cite{2danQCD}. In this model, the
analytic coupling $\A_1(Q^2)$ has no unphysical (Landau) singularities,
it predicts the correct measured value of $V$+$A$ strangeless
decay ratio of $\tau$ lepton,
and at high squared momenta it differs from the underlying perturbative
coupling $a(Q^2)$ ($\equiv \alpha_s(Q^2)/\pi$) by terms $\sim (\Lambda^2/Q^2)^5$,
allowing us to keep the ITEP school interpretation of the OPE expansion
up to (and including) dimension-eight ($D=8$) terms. The analysis
involves the evaluation of an integral of the $D=0$ Adler function 
${\cal D}(Q^2;D\!=\!0)$ along the contour $|Q^2|=m_{\tau}^2$. We evaluate this
function by applying a generalization of the diagonal Pad\'e resummation,
the method being exactly invariant under variation of the
renormalization scale and very convenient
for application within analytic QCD framework (the latter because no
Landau singularities appear). By comparing the experimental
ALEPH data of 1998 with the theoretical evaluations for the Borel
transform along the argument ray $M^2 = |M|^2 \exp(i \psi)$ with $\psi=\pi/6$, 
we extract with the standard
minimization for the ($D=4$) gluon condensate the values 
$\langle a G G \rangle = (0.0055 \pm 0.0047) \ {\rm GeV}^4$, cf.~Fig.~\ref{FigaGG} and
Eq.~(\ref{aGGres}). 
Furthermore, considering the ray $\psi= \pi/4$, we analogously extract for
$D=6$ condensate the
values $\langle O_6^{(V+A)} \rangle = (-0.5 \pm 1.1) \times 10^{-3} \ {\rm GeV}^6$,
cf.~Fig.~\ref{Figaqq} and Eqs.~(\ref{O6res})-(\ref{aqqres2}).
This is not incompatible with the theoretical expectation 
$\langle O_6^{(V+A)} \rangle \propto a \langle {\overline q} q \rangle^2 \geq 0$
based on the factorization (vacuum saturation)
approximation.
Using the obtained central values of the two condensates, 
$\langle a G G \rangle = 0.0055 \ {\rm GeV}^4$ and 
$\langle O_6^{(V+A)} \rangle = -0.5  \times 10^{-3} \ {\rm GeV}^6$,
the Borel transform on the real positive axis ($\psi=0$) gives the theoretical curve
which remains within the experimental band in the entire 
analyzed $M^2$-interval $(0.68 \  {\rm GeV}^2 < M^2 < 1.5 \ {\rm GeV}^2)$,
cf.~Fig.~\ref{Figpsi0}.

We also compare these results of the resummed analytic QCD approach
with those of the not resummed analytic QCD
approach and with those of perturbative QCD (pQCD) approaches 
in two schemes: Lambert scheme of the aformentioned analytic QCD model 
($c_2 \equiv \beta_2/\beta_0 = - 4.76$, $c_3=c_2^2/c_1$, etc.), and in 
$\MSbar$ scheme ($c_2 \approx 4.47$, $c_3 \approx 20.99$).
It turns out that in the Lambert scheme pQCD resummed approach and in the
$\MSbar$ pQCD nonresummed approach
the extracted central values of the gluon condensate $\langle a G G \rangle$
are similar, $0.0056$ and $0.0059 \ {\rm GeV}^4$, respectively,
and the $\chi^2$ values are only somewhat higher (by about $15 \%$).
On the other hand, in the not resummed pQCD approach in Lambert scheme,
and in the not resummed analytic QCD approach, 
the extracted values of the gluon condensate are significantly 
different, with the central value of about $0.0122$ 
and $0.0104 \ {\rm GeV}^4$, respectively, and $\chi^2$ values are by about
$50 \%$ higher than in the resummed analytic QCD approach.
On the other side, the values of $D=6$ condensate 
extracted (with $\psi=\pi/4$) in these approaches 
are negative, $\langle O_6^{(V+A)} \rangle \approx (-2 \pm 1) \times  10^{-3} \ {\rm GeV}^6$, 
suggesting a complete failure of the vacuum saturation approximation
for $\langle O_6^{(V+A)} \rangle$
(cf.~also Refs.~\cite{O6sr1,O6sr2} on this point). 
Furthermore, these approaches give us, for their corresponding
central values of $D=4$ and $D=6$ condensates, 
at $\psi=0$ the results which
are situated close to the lower edge of the experimental band
for all considered $M^2$.

Our result for the gluon condensate, $\langle a G G \rangle = (0.0055 \pm 0.0047) \ {\rm GeV}^4$,
is similar to the results of Refs.~\cite{Geshkenbein,Ioffe,O6sr1}. In Ref.~\cite{Geshkenbein},
the value $(0.006 \pm 0.012) \ {\rm GeV}^4$ was obtained from the $V$+$A$ channel of $\tau$-decay data,
and in Ref.~\cite{Ioffe} the value  $(0.005 \pm 0.004) \ {\rm GeV}^4$ from combination of the latter
(with ALEPH 1998 data) and of the charmonium sum rules. In Ref.~\cite{O6sr1}, where
ALEPH 2005 data were used, the approach of weighted finite energy sum rules
with the contour-improved perturbation theory gave
approximately the values $(0.008 \pm 0.005)  \ {\rm GeV}^4$ for the gluon condensate
when the standard $\MSbar$ QCD scale was\footnote{
This corresponds to $\alpha_s^{({\overline {\rm MS}})}(m_{\tau}^2;n_f=3)$ between
$0.2986$ and $0.3259$.} 
${\overline {\Lambda}}_{n_f=3} = 0.30$-$0.35$ GeV,
the higher values when ${\overline {\Lambda}} =0.3$ GeV 
and the lower values when ${\overline {\Lambda}} =0.35$ GeV. 
The original work on the sum rules,
Ref.~\cite{Shifman:1978bx}, obtains a higher estimate,
of $0.012  \ {\rm GeV}^4$ from charmonium physics.
QCD-moment and QCD-exponential moment sum rules for heavy quarkonia
give higher values,  $\langle a G G \rangle \approx (0.022 \pm 0.004) \ {\rm GeV}^4$
and $(0.024 \pm 0.006) \ {\rm GeV}^4$, respectively, Refs.~\cite{O4sr1}.
On the other hand, in Ref.~\cite{DDHMZ} a combined fit to the $V$+$A$ $\tau$-decay data
primarily extracts the value of $\alpha_s$ and, as a byproduct, obtains the
gluon condensate value  $\langle a G G \rangle = (-0.015 \pm 0.003) \ {\rm GeV}^4$ which differs
significantly from the aforementioned values.

The interpretation of our results can be summarized in the following. 
The nonexistence of the unphysical (Landau) singularities in the 
running coupling of the (2-delta-parametrized) 
analytic QCD framework of Ref.~\cite{2danQCD} 
leads to a reasonable degree of consistency between the theory and the
experiment in the Borel sum rule analysis of the strangeless
$V$+$A$ channel of $\tau$-decay physics. To achieve this, it is crucial
to perform, in addition, an efficient Pad\'e-related and
renormalization scale independent resummation of the Adler function,
a method which is very convenient for applications within analytic
QCD frameworks precisely because of the absence of the Landau
singularities.

\begin{acknowledgments}
\noindent
This work was supported in part by FONDECYT (Chile) Grants No.~1095196
(G.C.) and No.~1095217 (C.V.), and Rings Project ACT119 (G.C.).
\end{acknowledgments}

\appendix

\section{Renormalization scale independence of parameters in the resummed expression}
\label{app}

In this Appendix we show that the dimensionless parameters $\tal_j$ and
$\kappa_j$ ($j=1,2$), obtained via the construction in 
Eqs.~(\ref{tDpt})-(\ref{tQs}), are exactly independent of the 
(dimensionless) renormalization scale parameter $\kappa$ ($ \equiv \mu^2/Q^2$).

We recall that the truncated series ${\widetilde {\cal D}}(Q^2;\kappa)_{\rm pt}^{[4]}$ 
of Eq.~(\ref{tDpt}) is made of $\td_n(\kappa)$ coefficients which are
constructed from the coefficients $d_k(\kappa)$ of the original power
series (\ref{Dtpt}) via the relations (\ref{td1td2})-(\ref{td3}),
and whose $\kappa$ dependence is a consequence of $\kappa$ independence
of the full perturbation series ${\cal D}(Q^2)_{\rm pt}$ and
${\cal D}(Q^2)_{\rm mpt}$ of Adler function ${\cal D}(Q^2)$
(or any spacelike observable) [Eqs.~(\ref{Dpt})-(\ref{Dmpt})].
Further, the full series 
${\widetilde {\cal D}}(Q^2) = a_{1\ell}(\kappa Q^2) + 
\sum_{n=1}^{\infty} {\td}_n(\kappa) \; a_{1\ell}(\kappa Q^2)^{n+1}$ is $\kappa$-independent,
because of the differential relations (\ref{tdndiff}).

Let us choose two different renormalization scale parameters, $\kappa$ and
$\kappa^{\prime}$, and define the following notations
\bea
x & \equiv & a_{1 \ell}(\kappa Q^2) \ , \qquad
x^{\prime} \equiv a_{1 \ell}(\kappa^{\prime} Q^2) \ ,
\label{xxp}
\\
u(\kappa)  & \equiv & \beta_0 \ln \kappa \ .
\label{ukap}
\eea
The running of the coupling $a_{1 \ell}$ is one-loop, Eq.~(\ref{a1l}).
With the notation (\ref{ukap}), the quantities $\tu_j(\kappa)$'s appearing 
in the construction in Eq.~(\ref{dPA2tD}) [see also Eq.~(\ref{tQs})]
can be written as
\be
\tu_j(\kappa) = u(\kappa_j/\kappa) \quad \left( = \beta_0 \ln(\kappa_j/\kappa)
\right) \ .
\label{tujeq}
\ee
It is straightforward to verify from Eq.~(\ref{a1l}) for the one-loop running
coupling that $x$ and $x^{\prime}$ are related by
\be
x^{\prime} = \frac{x}{1 + u(\kappa^{\prime}/\kappa) x} \ ,
\label{xxprel}
\ee
and that the simple fractions appearing in Eq.~(\ref{dPA2tD}) can 
be reexpressed as
\bea
\frac{x}{1 + \tu_j(\kappa) x} \; \left( \equiv
\frac{x}{1 + u(\kappa_j/\kappa) x} \right)
& = & \frac{x^{\prime}}{1 + u(\kappa_j/\kappa^{\prime}) x^{\prime}} \ .
\label{sf1}
\eea
Further, it is straightforward to verify that the truncated
series ${\widetilde {\cal D}}(Q^2)_{\rm pt}^{[4]}$ of Eq.~(\ref{tDpt}),
while being $\kappa$-dependent due to truncation, at the two
renormalization scale parameters $\kappa$ and $\kappa^{\prime}$ differ from
each other by $\sim x^5$ ($\sim x^{\prime 5}$) terms
\be
P_{\kappa^{\prime}}(x^{\prime}) -  P_{\kappa}(x) 
\sim x^{\prime 5} \; (\sim x^5) \ ,
\label{PkpPk}
\ee
where the two truncated series of Eq.~(\ref{tDpt}) with $\kappa$ and
$\kappa^{\prime}$ are denoted as $P_{\kappa}(x)$ and $P_{\kappa^{\prime}}(x^{\prime})$ which are
quartic polynomials in $x$ and $x^{\prime}$, respectively
\bea
 P_{\kappa}(x) & \equiv & {\widetilde {\cal D}}(Q^2; \kappa)_{\rm pt}^{[4]}
= x + \sum_{n=1}^3 \td_n(\kappa) x^{n+1} \ ,
\label{Pk}
\\
 P_{\kappa^{\prime}}(x^{\prime}) & \equiv & {\widetilde {\cal D}}(Q^2; \kappa^{\prime})_{\rm pt}^{[4]}
= x^{\prime} + \sum_{n=1}^3 \td_n(\kappa^{\prime}) x^{\prime n+1} \ .
\label{Pkp}
\eea
Namely, the relation (\ref{PkpPk}) follows from the fact that
both $P_{\kappa}(x)$ and $P_{\kappa^{\prime}}(x^{\prime})$ differ from the
full $\kappa$-independent series ${\widetilde {\cal D}}(Q^2) = a_{1\ell}(\kappa Q^2) + 
\sum_{n=1}^{\infty} {\td}_n(\kappa) \; a_{1\ell}(\kappa Q^2)^{n+1}$ by terms 
$\sim x^5$ ($\sim x^{\prime 5}$).

On the other hand, Eq.~(\ref{sf1}) implies for the sum of simple
fractions appearing in the construction of the method, Eq.~(\ref{dPA2tD}),
the following identity:
\be
\sum_{j=1}^2 \tal_j \frac{x}{1 + \tu_j(\kappa) x} =
\sum_{j=1}^2 \tal_j \frac{x^{\prime}}{1 + u(\kappa_j/\kappa^{\prime}) x^{\prime}} \ .
\label{sf2}
\ee
The left-hand side of Eq.~(\ref{sf2}) is, by construction, the
diagonal Pad\'e $[2/2]_{P_{\kappa}}(x)$ of the polynomial
$P_{\kappa}(x)$ [$\equiv  {\widetilde {\cal D}}(Q^2; \kappa)_{\rm pt}^{[4]}$],
cf.~Eqs.~(\ref{dPA1tD})-(\ref{dPA2tD}).\footnote{
Note that any $[2/2](x)$ Pad\'e, being a ratio of two quadratic polynomials
with real coefficients, can always be decomposed into a sum of two simple
fractions of the form as on the left-hand side of Eq.~(\ref{sf2}),
where the coefficient pairs $\tal_j$ ($j=1,2$) and $\tu_j(\kappa)$ ($j=1,2$)
can now be complex conjugate pairs. In the case of Adler function
with $n_f=3$, it turns out that these pairs are real.} 
Therefore, Eq.~(\ref{sf2}) in conjunction with Eq.~(\ref{PkpPk}) implies that
the right-hand side of Eq.~(\ref{sf2}) is the diagonal Pad\'e
$[2/2]_{P_{\kappa^{\prime}}}(x^{\prime})$ of the polynomial $P_{\kappa^{\prime}}(x^{\prime})$ 
[$\equiv  {\widetilde {\cal D}}(Q^2; \kappa^{\prime})_{\rm pt}^{[4]}$]. More explicitly,
we have
\bea
\sum_{j=1}^2 \tal_j \frac{x^{\prime}}{1 + u(\kappa_j/\kappa^{\prime}) x^{\prime}} - P_{\kappa^{\prime}}(x^{\prime}) &=& 
\left( \sum_{j=1}^2 \tal_j \frac{x}{1 + \tu_j(\kappa) x} -  P_{\kappa}(x)
\right) + \left(  P_{\kappa}(x) - P_{\kappa^{\prime}}(x^{\prime}) \right)
\sim x^5 \sim x^{\prime 5} \ .
\label{PadxpPp}
\eea
The difference in the first pair of parentheses on the right-hand side
of Eq.~(\ref{PadxpPp}) is $\sim x^5$ because the sum of simple fractions there
is, by construction, the diagonal Pad\'e $[2/2]_{P_{\kappa}}(x)$ of the polynomial
$P_{\kappa}(x)$; the difference in the second pair of parentheses is
$\sim x^5$ due to the relation (\ref{PkpPk}). Eq.~(\ref{PadxpPp}) means
that the sum of the simple fractions on the left-hand side there is
the diagonal Pad\'e $[2/2]_{P_{\kappa^{\prime}}}(x^{\prime})$ 
of the polynomial $P_{\kappa^{\prime}}(x^{\prime})$.

However,
since  $P_{\kappa^{\prime}}(x^{\prime})$ is a polynomial whose coefficients are 
entirely independent of $\kappa$ (they are only $\kappa^{\prime}$-dependent),
this leads to the conclusion that the coefficients of the
Pad\'e $[2/2]_{P_{\kappa^{\prime}}}(x^{\prime})$ appearing on the right-hand side of
Eq.~(\ref{sf2}) are $\kappa$-independent, i.e.,
that $\tal_j$ and $u_j(\kappa_j/\kappa^{\prime})$ are $\kappa$-independent.
This means that $\kappa_j$ and $\tal_j$, which were obtained
via the construction in Eqs.~(\ref{tDpt})-(\ref{tQs}), 
are $\kappa$-independent. This concludes the demonstration.

The above argument can be almost literally repeated for the general
case of $N=2 M$ terms ($M=1,2,3,\ldots$) in the original truncated series 
(\ref{Dtpt}) and thus in the related truncated series
${\widetilde {\cal D}}(Q^2; \kappa^{\prime})_{\rm pt}^{[2 M]}$. The encountered
diagonal Pad\'e's are now $[M/M]$, and the decomposition in simple 
fractions is a sum of $M$ simple fractions.


\begin{thebibliography}{99}


\bibitem{BS}
N.N.~Bogoliubov and D.V.~Shirkov, 
{\it Introduction to the Theory of Quantum Fields\/}, New York, Wiley, 1959; 1980.

\bibitem{Oehme}
  R.~Oehme,
  Int.\ J.\ Mod.\ Phys.\  A {\bf 10}, 1995 (1995)
  [arXiv:hep-th/9412040].

\bibitem{ShS} 
  D.~V.~Shirkov and I.~L.~Solovtsov,
  hep-ph/9604363;
  Phys.~Rev.~Lett.~{\bf 79}, 1209 (1997)
[arXiv:hep-ph/9704333].

\bibitem{MSS}
  K.~A.~Milton, I.~L.~Solovtsov and O.~P.~Solovtsova,
  Phys.\ Lett.\ B {\bf 415}, 104 (1997)
[arXiv:hep-ph/9706409].

\bibitem{Sh}
  D.~V.~Shirkov,
  Theor.\ Math.\ Phys.\  {\bf 127}, 409 (2001)
[hep-ph/0012283];
  Eur.\ Phys.\ J.\ C {\bf 22}, 331 (2001)
  [hep-ph/0107282].


\bibitem{MSSY}
  K.~A.~Milton, I.~L.~Solovtsov, O.~P.~Solovtsova and V.~I.~Yasnov,
  Eur.\ Phys.\ J.\ C {\bf 14}, 495 (2000)
  [arXiv:hep-ph/0003030].

\bibitem{Milton:2001mq}
  K.~A.~Milton, I.~L.~Solovtsov and O.~P.~Solovtsova,
  Phys.\ Rev.\  D {\bf 64}, 016005 (2001)
  [arXiv:hep-ph/0102254].

\bibitem{mes2}
  M.~Baldicchi, A.~V.~Nesterenko, G.~M.~Prosperi, D.~V.~Shirkov and C.~Simolo,
  Phys.\ Rev.\ Lett.\  {\bf 99}, 242001 (2007);
M.~Baldicchi, A.~V.~Nesterenko, G.~M.~Prosperi, and C.~Simolo,
  Phys.\ Rev.\  D {\bf 77}, 034013 (2008).

\bibitem{Pase}
R.~S.~Pasechnik, D.~V.~Shirkov and O.~V.~Teryaev,
  Phys.\ Rev.\  D {\bf 78}, 071902 (2008)
  [arXiv:0808.0066 [hep-ph]];
R.~S.~Pasechnik, D.~V.~Shirkov, O.~V.~Teryaev, O.~P.~Solovtsova and V.~L.~Khandramai,
  Phys.\ Rev.\  D {\bf 81}, 016010 (2010)
  [arXiv:0911.3297];
  V.~L.~Khandramai, R.~S.~Pasechnik, D.~V.~Shirkov, O.~P.~Solovtsova and O.~V.~Teryaev,
  Phys.\ Lett.\ B {\bf 706}, 340 (2012)
  [arXiv:1106.6352 [hep-ph]].

\bibitem{NesSim}
  A.~V.~Nesterenko and C.~Simolo,
  Comput.\ Phys.\ Commun.\  {\bf 181}, 1769 (2010)
  [arXiv:1001.0901 [hep-ph]];
  Comput.\ Phys.\ Commun.\  {\bf 182}, 2303 (2011)
  [arXiv:1107.1045 [hep-ph]].

\bibitem{Bdecays}
  U.~Aglietti, G.~Ferrera and G.~Ricciardi,
  Nucl.\ Phys.\ B {\bf 768}, 85 (2007)
  [hep-ph/0608047];
  U.~Aglietti, F.~Di Lodovico, G.~Ferrera and G.~Ricciardi,
  Eur.\ Phys.\ J.\ C {\bf 59}, 831 (2009)
  [arXiv:0711.0860 [hep-ph]].

\bibitem{rev1}
  G.~M.~Prosperi, M.~Raciti and C.~Simolo,
  Prog.\ Part.\ Nucl.\ Phys.\  {\bf 58}, 387 (2007)
  [arXiv:hep-ph/0607209].

\bibitem{rev2}
  D.~V.~Shirkov and I.~L.~Solovtsov,
  Theor.\ Math.\ Phys.\  {\bf 150}, 132 (2007)
  [arXiv:hep-ph/0611229].

\bibitem{rev3}
  A.~P.~Bakulev,
  Phys.\ Part.\ Nucl.\  {\bf 40}, 715 (2009)
  [arXiv:0805.0829 [hep-ph]] (arXiv preprint in Russian).

\bibitem{Nest1}
  A.~V.~Nesterenko,
  Phys.\ Rev.\ D {\bf 62}, 094028 (2000);
  Phys.\ Rev.\ D {\bf 64}, 116009 (2001);
  Int.\ J.\ Mod.\ Phys.\ A {\bf 18}, 5475 (2003);

\bibitem{Nest2}
 A.~V.~Nesterenko and J.~Papavassiliou,
  Phys.\ Rev.\ D {\bf 71}, 016009 (2005);
  A.~C.~Aguilar, A.~V.~Nesterenko and J.~Papavassiliou,
  J.\ Phys.\ G {\bf 31}, 997 (2005)
  [hep-ph/0504195];
  Y.~.O.~Belyakova and A.~V.~Nesterenko,
  Int.\ J.\ Mod.\ Phys.\ A {\bf 26}, 981 (2011)
  [arXiv:1011.1148 [hep-ph]].

\bibitem{ALEPH1}
  R.~Barate {\it et al.}  [ALEPH Collaboration],
  Eur.\ Phys.\ J.\ C {\bf 4}, 409 (1998).

\bibitem{ALEPH2}
  S.~Schael {\it et al.}  [ALEPH Collaboration],
  Phys.\ Rept.\  {\bf 421}, 191 (2005)
  [hep-ex/0506072];
  M.~Davier, A.~H\"ocker and Z.~Zhang,
  Rev.\ Mod.\ Phys.\  {\bf 78}, 1043 (2006)
  [hep-ph/0507078].

\bibitem{DDHMZ}
  M.~Davier, S.~Descotes-Genon, A.~H\"ocker, B.~Malaescu and Z.~Zhang,
  Eur.\ Phys.\ J.\  C {\bf 56}, 305 (2008)
  [arXiv:0803.0979 [hep-ph]].


\bibitem{Shifman:1978bx}
  M.~A.~Shifman, A.~I.~Vainshtein and V.~I.~Zakharov,
  Nucl.\ Phys.\  B {\bf 147}, 385 (1979);
  Nucl.\ Phys.\  B {\bf 147}, 448 (1979).

\bibitem{DMW}
  Y.~L.~Dokshitzer, G.~Marchesini and B.~R.~Webber,
  Nucl.\ Phys.\ B {\bf 469}, 93 (1996)
  [arXiv:hep-ph/9512336].

\bibitem{panQCD}
  G.~Cveti\v{c}, R.~K\"ogerler, C.~Valenzuela,
  J.\ Phys.\ G {\bf G37}, 075001 (2010)
  [arXiv:0912.2466 [hep-ph]];
  Phys.\ Rev.\  {\bf D82}, 114004 (2010)
  [arXiv:1006.4199 [hep-ph]].

\bibitem{1danQCD}
  C.~Contreras, G.~Cveti\v{c}, O.~Espinosa and H.~E.~Mart\'{\i}nez,
  Phys.\ Rev.\  D {\bf 82}, 074005 (2010)
  [arXiv:1006.5050].

\bibitem{2danQCD}
  C.~Ayala, C.~Contreras and G.~Cveti\v{c},
  Phys.\ Rev.\ D {\bf 85}, 114043 (2012)
  [arXiv:1203.6897 [hep-ph]].

\bibitem{Alekseev}
  A.~I.~Alekseev,
  Few Body Syst.\  {\bf 40}, 57 (2006)
  [arXiv:hep-ph/0503242].

\bibitem{Nest3}
  A.~V.~Nesterenko,
  arXiv:1209.0164 [hep-ph].

\bibitem{DeRafael}
  S.~Peris, M.~Perrottet and E.~de Rafael,
  JHEP {\bf 9805}, 011 (1998)
  [arXiv:hep-ph/9805442].

\bibitem{MagrDual}
  B.~A.~Magradze,
  arXiv:1005.2674 [Unknown].

\bibitem{Geshkenbein}
  B.~V.~Geshkenbein, B.~L.~Ioffe and K.~N.~Zyablyuk,
  Phys.\ Rev.\  D {\bf 64}, 093009 (2001)
  [arXiv:hep-ph/0104048].

\bibitem{Ioffe}
  B.~L.~Ioffe,
  Prog.\ Part.\ Nucl.\ Phys.\  {\bf 56}, 232 (2006)
  [arXiv:hep-ph/0502148].

\bibitem{BGA} 
  G.~Cveti\v{c} and R.~K\"ogerler,
  Phys.\ Rev.\ D {\bf 84}, 056005 (2011)
  [arXiv:1107.2902 [hep-ph]].

\bibitem{Ioffe:2000ns} 
  B.~L.~Ioffe and K.~N.~Zyablyuk,
  Nucl.\ Phys.\ A {\bf 687}, 437 (2001)
  [hep-ph/0010089].


\bibitem{GCCV}
  G.~Cveti\v{c} and C.~Valenzuela,
  J.\ Phys.\ G {\bf 32}, L27 (2006)
  [arXiv:hep-ph/0601050];
  Phys.\ Rev.\  D {\bf 74}, 114030 (2006)
  [arXiv:hep-ph/0608256].

\bibitem{GCAK} 
  G.~Cveti\v{c} and A.~V.~Kotikov,
J.\ Phys.\ G  {\bf 39}, 065005 (2012)
  [arXiv:1106.4275 [hep-ph]].

\bibitem{BMS}
  A.~P.~Bakulev, S.~V.~Mikhailov and N.~G.~Stefanis,
  Phys.\ Rev.\  D {\bf 72}, 074014 (2005)
  [Erratum-ibid.\  D {\bf 72}, 119908 (2005)]
  [arXiv:hep-ph/0506311];
  Phys.\ Rev.\  D {\bf 75}, 056005 (2007)
  [Erratum-ibid.\  D {\bf 77}, 079901 (2008)]
  [arXiv:hep-ph/0607040];
  JHEP {\bf 1006}, 085 (2010)
  [arXiv:1004.4125 [hep-ph]];
  A.~P.~Bakulev and I.~V.~Potapova,
  Nucl.\ Phys.\ Proc.\ Suppl.\  {\bf 219-220}, 193 (2011)
  [arXiv:1108.6300 [hep-ph]].

\bibitem{Peris} 
  S.~Peris,
  Phys.\ Rev.\  D {\bf 74}, 054013 (2006)
  [arXiv:hep-ph/0603190].

\bibitem{CM}
  G.~Cveti\v{c} and H.~E.~Mart\'{\i}nez,
  J.\ Phys.\ G {\bf 36}, 125006 (2009)
  [arXiv:0907.0033 [hep-ph]].

\bibitem{Pades} 
G. A. Baker and P. Graves-Morris, {\it Pad\'e
Approximants\/}, Encyclopedia of Mathematics and its Applications
(Cambridge University, Cambridge, England, 1996).

\bibitem{Math8}
MATHEMATICA 8.0.4, Wolfram Co.

\bibitem{Gardi:1998qr}
  E.~Gardi, G.~Grunberg and M.~Karliner,
  JHEP {\bf 9807}, 007 (1998)
  [hep-ph/9806462].

\bibitem{Kou}
  D.~S.~Kourashev,
  arXiv:hep-ph/9912410.

\bibitem{KuMa}
D.~S.~Kurashev and B.~A.~Magradze,
  Theor.\ Math.\ Phys.\  {\bf 135}, 531 (2003)
  [Teor.\ Mat.\ Fiz.\  {\bf 135}, 95 (2003)].

\bibitem{Magr2}
  B.~A.~Magradze,
  Few Body Syst.\  {\bf 40}, 71 (2006)
  [hep-ph/0512374].

\bibitem{GCIK}
  G.~Cveti\v{c} and I.~Kondrashuk,
  JHEP {\bf 1112}, 019 (2011)
  [arXiv:1110.2545 [hep-ph]].

\bibitem{PDG2010}
  K.~Nakamura {\it et al.}  [Particle Data Group],
  J.\ Phys.\ G {\bf 37}, 075021 (2010).

\bibitem{CKS}
  K.~G.~Chetyrkin, B.~A.~Kniehl and M.~Steinhauser,
  Phys.\ Rev.\ Lett.\  {\bf 79}, 2184 (1997)
  [arXiv:hep-ph/9706430].

\bibitem{Shirkov:2006nc}
  D.~V.~Shirkov,
  Nucl.\ Phys.\ Proc.\ Suppl.\  {\bf 162}, 33 (2006)
  [arXiv:hep-ph/0611048].

\bibitem{BGApQCD1}
  G.~Cveti\v{c},
  Nucl.\ Phys.\  B {\bf 517}, 506 (1998)
  [arXiv:hep-ph/9711406];
  Phys.\ Rev.\  D {\bf 57}, 3209 (1998)
  [arXiv:hep-ph/9711487].

\bibitem{BGApQCD2}
  G.~Cveti\v{c} and R.~K\"ogerler,
  Nucl.\ Phys.\  B {\bf 522}, 396 (1998)
  [arXiv:hep-ph/9802248].

\bibitem{GardiPA}
  E.~Gardi,
  Phys.\ Rev.\  D {\bf 56}, 68 (1997)
  [arXiv:hep-ph/9611453].

\bibitem{d1}
  K.~G.~Chetyrkin, A.~L.~Kataev and F.~V.~Tkachov,
  Phys.\ Lett.\ B {\bf 85}, 277 (1979);
  M.~Dine and J.~R.~Sapirstein,
  Phys.\ Rev.\ Lett.\  {\bf 43}, 668 (1979);
  W.~Celmaster and R.~J.~Gonsalves,
  Phys.\ Rev.\ Lett.\  {\bf 44}, 560 (1980).

\bibitem{d2}
  S.~G.~Gorishnii, A.~L.~Kataev and S.~A.~Larin,
  Phys.\ Lett.\ B {\bf 259}, 144 (1991);
  L.~R.~Surguladze and M.~A.~Samuel,
  Phys.\ Rev.\ Lett.\  {\bf 66}, 560 (1991)
  [Erratum-ibid.\  {\bf 66}, 2416 (1991)].

\bibitem{d3}
  P.~A.~Baikov, K.~G.~Chetyrkin and J.~H.~K\"uhn,
  Phys.\ Rev.\ Lett.\  {\bf 101}, 012002 (2008)
  [arXiv:0801.1821 [hep-ph]].

\bibitem{ECH}
  G.~Grunberg,
  Phys.\ Rev.\  D {\bf 29}, 2315 (1984).

\bibitem{varsumrules}
  M.~Davier, A.~H\"ocker, L.~Girlanda, and J.~Stern,
  Phys.\ Rev.\ D {\bf 58} (1998) 096014
  [hep-ph/9802447];
  J.~Bijnens, E.~Gamiz and J.~Prades,
  JHEP {\bf 0110}, 009 (2001)
  [hep-ph/0108240];
  C.~A.~Dominguez and K.~Schilcher,
  Phys.\ Lett.\ B {\bf 581}, 193 (2004)
  [hep-ph/0309285];
  S.~Friot, D.~Greynat and E.~de Rafael,
  JHEP {\bf 0410}, 043 (2004)
  [hep-ph/0408281];
  S.~Narison,
  Phys.\ Lett.\ B {\bf 624}, 223 (2005)
  [hep-ph/0412152];
  J.~Bordes, C.~A.~Dominguez, J.~Penarrocha and K.~Schilcher,
  JHEP {\bf 0602}, 037 (2006)
  [hep-ph/0511293];
  S.~Bodenstein, J.~Bordes, C.~A.~Dominguez, J.~Penarrocha and K.~Schilcher,
  Phys.\ Rev.\ D {\bf 82}, 114013 (2010)
  [arXiv:1009.4325 [hep-ph]];
  S.~Bodenstein, C.~A.~Dominguez, S.~I.~Eidelman, H.~Spiesberger and K.~Schilcher,
  JHEP {\bf 1201}, 039 (2012)
  [arXiv:1110.2026 [hep-ph]];
and references therein.

\bibitem{O6sr1}
  C.~A.~Dominguez and K.~Schilcher,
  JHEP {\bf 0701}, 093 (2007)
  [hep-ph/0611347];

\bibitem{O6sr2}
  K.~Maltman and T.~Yavin,
  Phys.\ Rev.\ D {\bf 78}, 094020 (2008)
  [arXiv:0807.0650 [hep-ph]].


\bibitem{O4sr1}
  S.~Narison,
  Phys.\ Lett.\ B {\bf 706}, 412 (2012)
  [arXiv:1105.2922 [hep-ph]];
  Phys.\ Lett.\ B {\bf 707}, 259 (2012)
  [arXiv:1105.5070 [hep-ph]].

\bibitem{DV}
  D.~Boito, M.~Golterman, M.~Jamin, A.~Mahdavi, K.~Maltman, J.~Osborne and S.~Peris,
  Phys.\ Rev.\ D {\bf 85}, 093015 (2012)
  [arXiv:1203.3146 [hep-ph]];
and refences therein.

\end{thebibliography}
\end{document}